\documentclass[fleqn,usenatbib]{mnras}

\usepackage{newtxtext,newtxmath}

\usepackage[T1]{fontenc}

\DeclareRobustCommand{\VAN}[3]{#2}
\let\VANthebibliography\thebibliography
\def\thebibliography{\DeclareRobustCommand{\VAN}[3]{##3}\VANthebibliography}

\usepackage{graphicx}	
\usepackage{amsmath}	
\usepackage{CJKutf8}    
\usepackage{color}
\usepackage[dvipsnames]{xcolor}
\usepackage{comment}
\usepackage{bigstrut}
\usepackage[normalem]{ulem}

\newcommand{\lcdm}{{$\Lambda$CDM}}
\newcommand{\se}{{\sigma_8}}
\newcommand{\Om}{{\Omega_\mathrm{m}}}
\newcommand{\om}{{\omega_\mathrm{m}}}
\newcommand{\Se}{{S_8}}
\newcommand{\be}{\begin{equation}}
\newcommand{\ee}{\end{equation}}
\newcommand{\ba}{\begin{aligned}}
\newcommand{\ea}{\end{aligned}}
\newcommand{\bes}{\begin{equation*}}
\newcommand{\ees}{\end{equation*}}
\newcommand{\bea}{\begin{eqnarray}}
\newcommand{\eea}{\end{eqnarray}}
\newcommand{\dd}{\text{d}}
\newcommand{\eg}{\textit{e.g.}}

\font \bolditalics = cmmib10
\def\bx#1{\leavevmode\thinspace\hbox{\vrule\vtop{\vbox{\hrule\kern1pt
        \hbox{\vphantom{\tt/}\thinspace{\bf#1}\thinspace}}
      \kern1pt\hrule}\vrule}\thinspace}
\def \vc #1{{\textfont1=\bolditalics \hbox{$\bf#1$}}}  

\def\thetag{{\vc \theta}}

\newcommand{\chinared}[1]{{\color[HTML]{EE1C25}{{#1}}}}

\newcommand{\salmon}[1]{{\color[RGB]{250,128,114}{{#1}}}}
\newcommand{\cyan}[1]{{\color[RGB]{0,255,255}{{#1}}}}

\newcommand{\darkorange}[1]{{\color[RGB]{255,140,0}{{#1}}}}
\newcommand{\brightazure}[1]{{\color[RGB]{30,144,255}{{#1}}}}

\newcommand{\limegreen}[1]{{\color[RGB]{50,205,50}{#1}}}
\newcommand{\gold}[1]{{\color[RGB]{255,215,0}{#1}}}

\newcommand{\slategrey}[1]{{\color[RGB]{112,128,144}{{#1}}}}

\newcommand{\darkred}[1]{{\color[RGB]{139,0,0}{#1}}}
\newcommand{\indigo}[1]{{\color[RGB]{75,0,130}{#1}}}

\newcommand{\rosybrown}[1]{{\color[RGB]{188,143,143}{#1}}}

\newcommand{\olive}[1]{{\color[RGB]{128,128,0}{#1}}}
\newcommand{\lime}[1]{{\color[RGB]{0,255,0}{#1}}}
\newcommand{\orchid}[1]{{\color[RGB]{218,112,214}{#1}}}

\title[Weak Lensing constraints of the HMF and Cosmology]{On constraining Cosmology and the Halo Mass Function with Weak Gravitational Lensing.}

\author[Shiming Gu \begin{CJK*}{UTF8}{gbsn}(顾时铭)\end{CJK*} et al.]{
Shiming Gu \begin{CJK*}{UTF8}{gbsn}(顾时铭)\end{CJK*}$^{1}$\thanks{E-mail: gsm@phas.ubc.ca (SG)}
Marc-Antoine Dor,$^{1,2}$
Ludovic van Waerbeke,$^{1}$
Marika Asgari,$^{3,4,5}$ \newauthor
Alexander Mead,$^{6,8}$
Tilman Tr\"{o}ster,$^{5,7}$
and Ziang Yan\begin{CJK*}{UTF8}{gbsn} (颜子昂)\end{CJK*}$^{8}$
\\
$^{1}$Department of Physics and Astronomy, University of British
Columbia, Vancouver, BC, V6T 1Z1, Canada \\
$^{2}$ENSTA Paris, 828, Boulevard des Mar\'{e}chaux, 91762 Palaiseau Cedex, France \\
$^{3}$E.A Milne Centre for Astrophysics, University of Hull, Cottingham Road, Kingston upon Hull, HU6 7RX, UK \\
$^{4}$Centre of Excellence for Data Science, AI, and Modelling (DAIM), University of Hull, Cottingham Road, Kinston upon Hull, HU6 7RX \\
$^{5}$Institute for Astronomy, University of Edinburgh, Royal Observatory, Blackford Hill, Edinburgh, EH9 3HJ, UK \\
$^{6}$Department of Computer Science, University of British Columbia, Vancouver, BC V6T 1Z1, Canada \\
$^{7}$Eidgenoessische Technische Hochschule Z{\"u}rich,
R{\"a}mistrasse 101, 8092 Zurich, Switzerland \\
$^{8}$Ruhr University Bochum, Faculty of Physics and Astronomy, Astronomical Institute (AIRUB), German Centre for Cosmological Lensing, 44780 Bochum, Germany \\
}

\date{Accepted XXX. Received YYY; in original form ZZZ}

\pubyear{2023}

\begin{document}
\label{firstpage}
\pagerange{\pageref{firstpage}--\pageref{lastpage}}
\maketitle

\begin{abstract}
{The discrepancy between the weak lensing (WL) and the {\it Planck} measurements of $\Se$ has been a subject of several studies. Assuming that residual systematics are not the cause, these studies tend to show that a strong suppression of the amplitude of the mass power spectrum $P(k)$ in the late universe at high $k$ could resolve it. The WL signal at the small scale is sensitive to various effects not related to lensing, such as baryonic effects and intrinsic alignment. These effects are still poorly understood, therefore, the accuracy of $P(k)$ depends on the modelling precision of these effects. A common approach for calculating $P(k)$ relies on a halo model.
 Amongst the various components necessary for the construction of $P(k)$ in the halo model framework, the halo mass function (HMF) is an important one. Traditionally, the HMF has been assumed to follow a fixed model, motivated by dark matter-only numerical simulations. Recent literature shows that baryonic physics, amongst several other factors, could affect the HMF.
In this study, we investigate the impact of allowing the HMF to vary. This provides a way of testing the validity of the halo model-HMF calibration using data. In the context of the aforementioned $\Se$ discrepancy, we find that the {\it Planck} cosmology is not compatible with the vanilla HMF for both the DES-y3 and the KiDS-1000 data. Moreover, when the cosmology and the HMF parameters are allowed to vary, the {\it Planck} cosmology is no longer in tension. The modified HMF predicts a matter power spectrum with a $\sim 25\%$ power loss at $k\sim 1~{\rm h/Mpc}$, in agreement with the recent studies that try to mitigate the $\Se$ tension with modifications in $P(k)$. We show that Stage IV surveys will be able to measure the HMF parameters with a few percent accuracy.}
\end{abstract}




\section{Introduction}

The presence of dark matter in our Universe has been firmly established, with baryonic matter accounting for approximately 20\% of the total mass, while dark matter (DM) comprises around 80\% \citep[\eg][]{planck18,Aiola+20,eboss20}. Weak gravitational lensing is a well-known technique employed to investigate the mass distribution by analyzing the distorted shapes of background galaxies \citep{Gunn67,Blandford+91}. Over the past decade, several cosmic shear surveys have been conducted, including the Canada-France-Hawaii-Telescope Lensing Survey (CFHTLens) \citep[]{CFHTLS}, the Kilo Degree Survey (KiDS) \citep[]{Kuijken+15}, the Dark Energy Survey (DES) \citep[]{Drlica-Wagner+18}, and the Hyper Suprime-Cam (HSC) \citep[]{Aihara+18}.

The cosmological interpretation of weak lensing measurements necessitates the computation of the matter power spectrum $P(k)$, which is a function of redshift and cosmological parameters. However,  calculating $P(k)$ accurately is challenging, particularly at the small scale, where some contributions are not constrained well enough. These contributions arise from factors such as intrinsic alignment, baryonic feedback, and neutrino streaming, among others. Consequently, semi-analytical models of $P(k)$ are calibrated with numerical simulations \citep[\eg][]{PeacockSmith00,HALOFIT,Takahashi+12,Bird+12}.
The issue is that differences in the calibration recipes, employed by different models, lead to different $P(k)$, which consequently impact the cosmological parameter inference at the level of a few percents. While this problem has been recognized for quite some time, several modifications to the calibration procedure have been implemented. However, as future lensing surveys strive for higher accuracy, the calibration of $P(k)$ may once again become the main limiting factor, thereby preventing lensing surveys from reaching their full potential. The  $\Se$ tension observed between low and high redshifts\footnote{$\Se = \se\sqrt{\Om/0.3}$} could be an indication of this issue. If this tension is genuine, its underlying cause (residual systematics, modelling of $P(k)$, or new physics), remains unknown.

Recently, the joint analysis of the Dark Energy Survey (DES-y3) and the Kilo-Degree Survey (KiDS-1000) was conducted by \cite{DES+KiDS23}. They discovered that by swapping the analysis pipeline of each survey, the $\Se$ tension between the surveys increased significantly, hence exacerbating the tension between KiDS and {\it Planck} while decreasing it between DES and Planck. However, this issue was resolved by adopting a hybrid analysis pipeline that incorporates elements from both surveys.
The solution emerged from a tuning of the intrinsic alignment and baryonic effects models, tested on numerical simulations. Consequently, the $\Se$ tension with the cosmic microwave background diminished, while KiDS-1000 and DES-y3 exhibited good agreement. It should be noted that this does not provide a definitive solution for modelling $P(k)$, but rather emphasizes the challenges associated with developing a reliable and predictive high-precision model for $P(k)$.
In \cite{DES+KiDS23} and other papers aimed at mitigating the $\Se$ tension, the focus consistently revolves around the modelling of $P(k)$. For instance, \cite{Amon+22c} proposed a one-parameter modification of the matter power spectrum on a non-linear scale.\footnote{However, they demonstrated that achieving a reduction in the $\Se$ tension would necessitate an unrealistic suppression of power at the non-linear scale.}

Not all options impacting the modelling of $P(k)$ have been explored. For instance, semi-analytical models, such as the halo model based \textsc{HMcode} \citep[][]{Mead+15}, are widely used to calculate $P(k)$. At the heart of any halo model, there is a Halo Mass Function (HMF) \citep[\eg,][]{PSHMF,STHMF} a fundamental building block from which $P(k)$ is constructed. The parameters describing the HMF are set to fixed values, usually determined by DM only $N$-body simulations, and other modifications are supposed to take care of all the complexities of non-linear physics. The HMF is not an exact description of the haloes distribution in the universe, but it is a proxy that serves this purpose. As summarized below, there are good reasons for wanting the parameters of the HMF to be adjustable, which is the main motivation of our study.

A direct measurement of the HMF would significantly improve the modelling of $P(k)$. Unfortunately, this is not possible, and, at best, the HMF can be partially constrained using baryonic probes as tracers of DM haloes, which involves various baryon-based quantities. \citet{Castro+16} discussed the constraints of Sheth-Tormen HMF models \citep[]{STHMF,Despali+16} by analyzing observations of galaxy cluster counts, cluster power spectra, and Type Ia supernova lensing. One possibility is to use the size and velocity dispersion relations of galaxy clusters and/or groups \citep[\eg][]{BahcallCen93,Boehringer+17}, but this method relies on a simulation-calibrated relationship between mass and X-ray luminosities \citep[][]{Hoekstra+11}.
Previous work using $N$-body simulations has shown that modifications of the Halo Mass Function (HMF) can arise from distinct initial conditions and/or dark matter physics. For example, \citet{Bagla+09} demonstrated that the HMF changes depending on the spectral indices of the primordial power spectrum, while \citet{Costanzi+13} showed that the HMF's shape depends on the neutrino mass and matter density. \citet{Barreira+14} showed that the spherical collapse is changed under an alternate theory of gravity, thereby causing the HMF calibration from DM-only N-body simulations to be inaccurate. Using hydro simulations, \citet{Bocquet+16} showed that baryonic physics impacts the shape of the HMF, and \citet{Baugh+19} showed that the HMF varies by $25\%$ between the WMAP7-based and Planck-based mock catalogues due to diverse gas cooling rates triggered by different $\omega_\mathrm{b}$ values \citep{WFSAM}. The HMF also depends on the specifics of the halo-finding algorithm \citep{Knebe+11}. In summary, it was shown in various papers that the HMF is not universal that it depends on underlying physical processes, cosmology and selection effects \citep{Ondaro-Mallea+22,Diemer21,Castro+21}. 
Therefore, we believe that it is important to incorporate the HMF parameters as degrees of freedom in the modelling of $P(k)$ in order to better represent our ignorance and see if the fixed values of the HMF parameters are recovered by the fitting process. If they are not, this would indicate a problem with the data and/or the modelling. Our approach is inspired by the shape measurement process to test residual systematics, where a shear calibration parameter $m$ was introduced at the model fitting stage, which should be zero if no residual systematics are present in the galaxy shapes measurement; all lensing surveys generally find a small but non-zero value. Similarly, the free HMF parameters are introduced at the model fitting level in order to verify if the theoretically inferred values are acceptable.

This paper is structured as follows. In section \ref{secmd}, we provide an introduction to the halo mass function and explain how it affects weak lensing observables. In section \ref{secdt}, we discuss the modelling process and the data used in this study. Parameter interference from the observables is discussed in section \ref{secpi}. In section \ref{seccc}, we present the parameter constraint on the HMF and cosmology based on the new HMF. Finally, we summarize our findings and conclude in section \ref{seccl}.

\section{Theory}
\label{secmd}

\subsection{The Shear Correlation Function}

This study is based on the comparison between the measured shear correlation function and the theory, using the halo model approach and the DES-y3 and KiDS surveys. We decided to base our analysis on cosmic shear only and not include galaxy clustering nor galaxy-galaxy lensing because we want an estimator completely free of any assumption of galaxy bias, assembly bias, or any consideration regarding on how galaxies populate haloes.

The shear two-point correlation functions is $\xi_{\pm}^{\rm ij}(\theta)$, where $i$ and $j$ refer to redshift bins. The estimator $\hat\xi_{\pm}^{\rm ij}(\theta)$ is given by:

\begin{equation}
\hat\xi_{\pm}^\mathrm{ij}(\theta) = \frac{\sum_{a,b}w_\mathrm{a} w_\mathrm{b}\left[\epsilon^i_t(\thetag_a)\epsilon^j_\mathrm{t}(\thetag_\mathrm{b})\pm \epsilon^i_\times(\thetag_\mathrm{a})\epsilon^j_\times(\thetag_\mathrm{b})\right]}{\sum_{a,b}w_\mathrm{a} w_\mathrm{b}},
\end{equation}

where $\thetag_\mathrm{a}$ and $\thetag_\mathrm{b}$ are galaxies' position vectors and the summation is taken over galaxy pairs $(a,b)$. The galaxies have an angular separation $|\thetag_\mathrm{a}-\thetag_\mathrm{b}|$ within an interval $\Delta\theta$ centered on $\theta$. The galaxy weights are represented as $w_\mathrm{a}$ and $w_\mathrm{b}$, while $\epsilon_{\mathrm{t},\times}^{i,j}$ refers to the tangential and cross ellipticities of galaxies in redshift bins $i$ and $j$.

The estimator $\hat\xi_{\pm}^{\rm ij}(\theta)$ can be expressed in terms of the convergence power spectrum $C_{\kappa\kappa}^{ij}(\ell)$ \citep[][]{Hamilton00}:

\be \label{eqn:shear_calc_transform}
\xi_\pm^{ij}(\theta) = \frac{1}{2\pi}\int_{0}^{\infty}\dd\ell\ \ell J_{0,4}(\ell\theta)C_{\kappa\kappa}^{ij}(\ell),
\ee

Here, $J_{0,4}$ represents the 0- and 4-th orders of the Bessel functions of the first kind. For scales $\ell>10$, it is possible to approximate the sky as flat, and the angular cross-spectrum $C_{\kappa\kappa}^{ij}(\ell)$ is related to the 3D power spectrum $P_\delta$ using the Limber approximation \citep{Limber53}:

\be \label{eqn:limber1}
C_{\kappa\kappa}^{ij}(\ell) = \int_0^{\chi_{\rm H}}\dd\chi \frac{q_i(\chi)q_j(\chi)}{[f_K(\chi)]^2} P\left(k=\frac{\ell+\dfrac{1}{2}}{f_K(\chi)},\chi \right),
\ee

In the above equation, $\chi$ is the radial comoving distance, $f_K(\chi)$ is the angular diameter distance, $\chi_{\rm H}$ is the horizon distance, and $P(k,\chi)$ is the 3D matter power spectrum. The lensing efficiency function, for the redshift bin $i$, $q_i(\chi)$, is given by:

\begin{equation}
q_\mathrm{i}(\chi) = \frac{3H_0^2 \Om}{2c^2}\int_\chi^{\chi_{\rm H}}\frac{\dd \chi'}{a(\chi)} n_\mathrm{i}(\chi')\frac{f_K(\chi'-\chi)f_K(\chi)}{f_K(\chi')},
\end{equation}
where, $H_0$ is the Hubble Constant \citep[][]{Hubble36b}, $\Om$ the matter density fraction of the universe, $c$ the speed of light, $a(\chi)$ the dimensionless scale factor, and $n_\mathrm{i}(\chi)$ is the effective number of galaxies in bin $i$ normalized to $\int_0^{\chi_{\rm H}} \dd\chi n(\chi) = 1$.

The power spectrum $P(k)$ is the contribution of two parts. The first part is due to matter clustering within a halo and is called the one-halo term, $P_{\text{1H}}(k)$. The second part arises from clustering between haloes and is called the two-halo term, $P_{\text{2H}}(k)$:
\be
P(k) = P_{\text{1H}}(k) + P_{\text{2H}}(k).
\ee
In the following two Sections, we introduce the Halo Model in general terms and then describe the specific \textsc{HMcode} \citep[][]{Mead+15} implementation that we use to do the calculations.

\subsection{The Halo Profile and the Halo Mass Function}
\label{HMFtheory}

A recent review of the halo model can be found in \cite{Asgari+23}.
The theoretical ingredients for the halo model are the halo mass density profile $\rho(r,M)$ and the halo mass function (HMF) $f(\nu)$, where $\nu$ is the peak threshold of a halo of mass $M$, defined as:

\begin{equation}\label{eqn:nu}
\nu = \frac{\delta_c(z)}{\sigma(M,z)},
\end{equation}
where $\delta_c(z)$ is the linear-theory-based critical threshold for spherical collapse, and the variance $\sigma^2(M,z)$ is computed from the linear matter power spectrum $\Delta^2_{\mathrm{lin}}(k,z)$:

\begin{equation}
\sigma^2(M,z) = \int_0^\infty \dd k \ k^{-1}\Delta^2_{\mathrm{lin}}(k,z)W^2(k,r),
\end{equation}
where $W(k,r)$ is a Fourier transformed spherical top-hat function. The function $f(\nu)$ can be expressed in terms of the number density $n(M)$ of haloes of mass $M$:

\be\label{eqn:fnunM}
f(\nu) = n(M)\frac{\bar\rho}{M}\frac{\dd\nu}{\dd M},
\ee
where $\bar{\rho}$ is the mean matter density. The shape of $f(\nu)$ is determined by two parameters, $p$ and $q$, as described in \citet[][]{STHMF}. The parameter $p$ is associated with the tidal field between overlapping DM haloes, which determines the abundance of low-mass haloes. The parameter $q$ is associated with the boundary crossing between two haloes, which modifies the spherical collapse of the DM halo by correcting the peak threshold $\nu'^2 = q\nu^2$ \citep[][]{Sheth+01}. The function $f(\nu)$ and the normalisation constant  $A$ are given by:

\be\label{eqn:fnu}
f(\nu) = A\sqrt{
\frac{2q}{\pi}}\left(1+\left(\frac{1}{\nu'^2}\right)^{-p}\right)e^{-\frac{\nu'^2}{2}},
\ee
\be\label{eqn:Ap}
A(p) = \frac{1}{1+\pi^{-1/2}2^{-p}\Gamma(\frac{1}{2}-p)}.
\ee
Although $f(\nu)$ was originally constructed as a universal function that has no cosmological evolution or power spectrum shape at all, its form can still change drastically when baryonic feedback or exotic dark matter models are considered, as has been discussed in the introduction.
For instance, the parameter $q$, which scales with the peak threshold $\nu$, is relevant to baryonic feedback \citep[][]{VanDaalen+20} and neutrinos \citep[][]{Costanzi+13}, as both effects change the peak threshold for spherical collapse. On the other hand, $p$ determines the abundance gradient from low-mass to high-mass haloes. As shown by simulations of exotic dark matter models, this gradient is sensitive to specific dark matter properties \citep[\eg,][]{Lovell20,Kulkarni+22}. The best-fit values of $(p,q)$ given by \citet{STHMF} are $p=0.3$ and $q=0.707$, which yields $A = 0.322$. Another approach to the Sheth-Tormen mass function is applied by \citet{Despali+16}, where $A$ is set to be a free parameter, which can affect the vanilla two-halo term at large $k$ \citep[][]{Mead+20a}. In this paper, we will consider several mathematically normalized ST-like HMF models given by Equation \ref{eqn:Ap} (labelled ST-X), as well as a few models that are not normalized (labelled DG-X).

Finally, it is important to clarify the concept of the halo definition. The original ST model uses the spherical overdensity group finder, based on the virial mass definition \citep[][]{Tormen98}, which takes into account the evolution of halos over cosmic time \citep[][]{jenkins01}. This virial mass definition was also used in the calibration of \textsc{HMcode} \citep[][]{Mead+15}. Therefore, as our primary objective is to check the consistency of the WL analysis pipelines, we will adhere to the virial definition that has been used in conjunction with the ST model and \textsc{HMcode} calibration.

\subsection{The Halo Model and \textsc{HMcode}}

In this paper, we propose a modification of the matter power spectrum by allowing the Sheth-Tormen HMF parameters to vary. Our work has a similar motivation to \cite{Amon+22c}, which uses a had-oc modification of $P(k)$ at the small scale. \cite{Amon+22c} introduced the free parameter $A_{\rm mod}$ such that:

\begin{equation}
    P(k)=P^{\rm L}(k)+A_{\rm mod}[P^{\rm NL}-P^{\rm L}]
    \label{Amon}
\end{equation}
where $P^{\rm L}(k)$ and $P^{\rm NL}$ are the linear and non-linear power spectra respectively. Using the KiDS-1000 data, they found that $A_{\rm mod}=0.69\pm 0.1$ best fit the data when forcing the Planck18 cosmology. This corresponds to a power loss of the order of $\sim 30\%$ in the non-linear scales, in addition to the fiducial KiDS-1000 fiducial cosmology. The underlying causes for this power loss could be the superposition of many different effects. In our work, the proposed modification of $P(k)$ consists of an alteration of the HMF, which is justified by the fact that there are several clearly identified reasons for the HMF to be modified compared to its vanilla functional form, as discussed in the Introduction.

For our calculations, we will use \textsc{HMcode} \citep[][]{Mead+16}. In order to solve the problem of missing power at intermediate scales with the halo model, \textsc{HMcode} implements a modified one-halo $P'_{\text{1H}}$ and two-halo $P'_{\text{2H}}$ terms such that:
\be
P(k) = \left[(P'_{\text{1H}}(k))^{\alpha} + (P'_{\text{2H}}(k))^{\alpha}\right]^{\frac{1}{\alpha}},
\label{haloterms}
\ee
where $\alpha$ is a smoothing parameter that depends on the scale where the halo collapse occurs. The modified one-halo term takes into account the damping from the halo exclusion effect \citep{Smith+07}. Specifically, $P'_\mathrm{1H} = \left[1-e^{-(k/k_*)^2}\right] P_\mathrm{1H}$, with $k_*$ being the one-halo damping wavenumber given by \citet{Mead+16}. The two-halo term is damped at quasi-linear scales following the calculation of \citet{CrocceScoccimarro06}, which means that only $P'_{\text{1H}}$ in \textsc{HMcode} depends on $p$ and $q$, not the 2-halo term (unlike the vanilla halo model).

Within the halo model, the one-halo term is given by:

\begin{equation} \label{eqn:1ht}
P_{\text{1H}}(k) = \frac{1}{\bar\rho} \int_0^\infty \dd\nu\ M(\nu)\tilde\rho^2(\nu^\eta k,M)f(\nu),
\end{equation}
where, $M$ represents the halo mass and $\eta$ is the halo bloating parameter. The normalized Fourier transform of the halo density profile is given by $\tilde\rho^2(k,M)$, which was calibrated to the Navarro-Frenk-White (NFW) profile. The NFW profile is defined by the equation:

\begin{equation}
\rho(r) \propto \frac{r_\mathrm{s}}{r}\left(1+\frac{r}{r_\mathrm{s}}\right)^{-1},
\end{equation}

where $r_\mathrm{s}=r_\mathrm{virial}/c(M,z)$ is the scale radius and $c(M,z)$ is the halo concentration. The \textsc{HMcode} uses the halo concentration recipe developed by Bullock et al. (2001). It is worth noting that the halo concentration relation can have a significant impact on small scales and may have an internal degeneracy with HMF parameters. However, in the \textsc{HMcode} framework, changing the halo concentration recipe can result in a worse $P(k)$ fit to simulations compared to just changing the HMF recipe, something which has already been tested in the \textsc{HMcode}. Therefore, we keep the original Bullock et al. (2001) halo concentration relation in the \textsc{HMcode}.

\be
c(M,z) = B\frac{1+z_\mathrm{HF}(M)}{1+z},
\ee

Here, the halo-forming redshift $z_\mathrm{HF}(M)$ can be found in Bullock et al. (2001). The parameter $B$ represents the minimum halo concentration and is a useful indicator for baryonic feedback. A smaller value of $B$ corresponds to a stronger cooling or a larger AGN feedback \citep{VanDaalen+20,Mead+21}. This parameter was added to the KiDS-VIKING-450 \citep[][]{Hildebrandt+20} and KiDS-1000 pipelines \citep[][]{Asgari+21} with a flat prior and a very limited range.

\section{Data and Likelihood}

Two lensing surveys are used in this work: the Kilo-Degree Survey fourth data release \citep[KiDS-1000][]{Kuijken+19,Giblin+21} and the Dark Energy Survey year 3 data release \citep[DES-y3][]{DES+22}. In addition to those two datasets, we also use the LSST mock lightcone from the Scinet Light Cone Simulations \citep[SLICS][]{Harnois-Deraps+18} to provide comparisons between the current state-III results and the future Stage-IV surveys.

\subsection{KiDS}

The Kilo-Degree Survey fourth data release dataset (KiDS-1000) covers an unmasked field of view of $1006$ deg$^2$. This survey is based on data obtained from two telescopes, both of which are part of the \textit{Very-Large Telescope} (VLT). The main survey telescope, the $2.65$m \textit{VLT Survey Telescope} (VST), provides $4$ optical bands ($ugri$) with the \textit{VST-OmegaCAM} \citep[][]{Kuijken11}, which has a mean seeing of $0.7''$ in the $r$-band. Additionally, the $4.1$m \textit{Visible and Infrared Survey Telescope for Astronomy} (VISTA) provides another $5$ near-infrared bands ($ZYJHK_s$) through the \textit{VISTA Kilo-degree Infrared Galaxy survey} \citep[VIKING,][]{Edge+13}, which were introduced in the updated sample KiDS-VIKING-450 of the third release of the same survey, KiDS-450 \citep[][]{deJong+17,Hildebrandt+17}. The VIKING infrared bands reach depths of $r \leq 25$ and cover redshifts up to $z \leq 6$.
 
The KiDS galaxy catalogue is divided into five photometric-redshift bins. The photometric-redshifts used in KiDS-1000 were obtained by applying the Bayesian photometric-redshift (BPZ) technique \citep[][]{Benitez00,Raichoor+14}, and were refined with the self-organising map (SOM) method \citep[SOM,][]{SOM,Geach12,Wright+20b}, using spectroscopic data for calibration. The redshift distribution for each bin was determined based on the spectroscopic redshifts of similar galaxies. Any galaxy that could not be grouped with spectroscopic galaxies were removed from the '\textsc{Gold}' catalogue. The accuracy of these redshift distributions was confirmed with the mock data from the MICE2 simulation \citep[][]{Fosalba+15a,Fosalba+15b,Crocce+15} \citep[][]{vandenBusch+22}. The fiducial data products and corresponding covariance matrices used in our study are taken from \citet{Asgari+21,Giblin+21} and \citet{Joachimi+20}, respectively, as official KiDS-1000 samples.
 
\subsection{DES}

The Dark Energy Survey Year 3 data release (DES-y3) is shallower and wider than KiDS-1000, covering $5000$ deg$^2$ of the sky. DES employs the Dark Energy Camera (DECam) \citep[DECam][]{Flaugher+15}, which is mounted on the 4m Victor M. Blanco Telescope located at Cerro-Tololo Observatory. The survey comprises of five bands: four optical bands ($griz$) and one infrared band ($Y$), although, effectively only three bands (riz) are used in the shapes and photometric redshift estimates. The survey's depth is $i \leq 24$, which roughly corresponds to $r \leq 24.4$. The average seeing is approximately $1'$ in the $r$-band \citep[][]{Sevilla-Noarbe+21}.

For our tomographic re-analysis of DES-y3, we use the same four redshift bins and data vector as in DES-y3 cosmic shear \citep{Amon+22a,Secco+22}. The DES photometric redshifts are determined using the calibration procedure outlined in \citet{Myles+21}. This procedure involves three estimates for $n(z)$: Bayesian probability distributions obtained from the self-organizing map (SOMPZ) \citep[][]{Hartley+22}, whose Gaussian prior is derived from the combination of distance constraints from clustering redshifts \citep[][]{Newman08,Gatti+22} and shear ratios \citep[][]{JainTaylor03,Sanchez+22}. All three methods are independent of each other. We use the covariance matrices from the fiducial DES-y3 \citet{Fang+20} analyses based on the CosmoLike framework \citep[][]{KrauseEifler17}.

\subsection{SLICS LSST Mock Lightcone}
Scinet Light Cone Simulations (SLICS) \citep[][]{HDvW15} is a large set of simulations designed for data quality assessment and covariance estimation of multiple weak lensing data analyses, including the KiDS-1000 and LSST. It is based on a series of $1025$ N-body simulation boxes using the best-fit of WMAP9 + BAO + SN parameters in \citet{WMAP9} ($\Omega_m = 0.2905$, $\Omega_b = 0.0473$, $h = 0.6898$, $\sigma_8 = 0.826$, $n_s = 0.969$). In order to perform the mock analysis of future Stage-IV surveys of LSST, we use all $819$ LSST-like Mock lightcones from SLICS \citep[][]{Harnois-Deraps+18} to construct our covariance matrices for all 10 LSST-like tomographic bins, as well as the corresponding 10-bin tomographic redshift distributions.

\subsection{Likelihood}

\begin{figure*}
\centering
\includegraphics[width=0.45\textwidth]{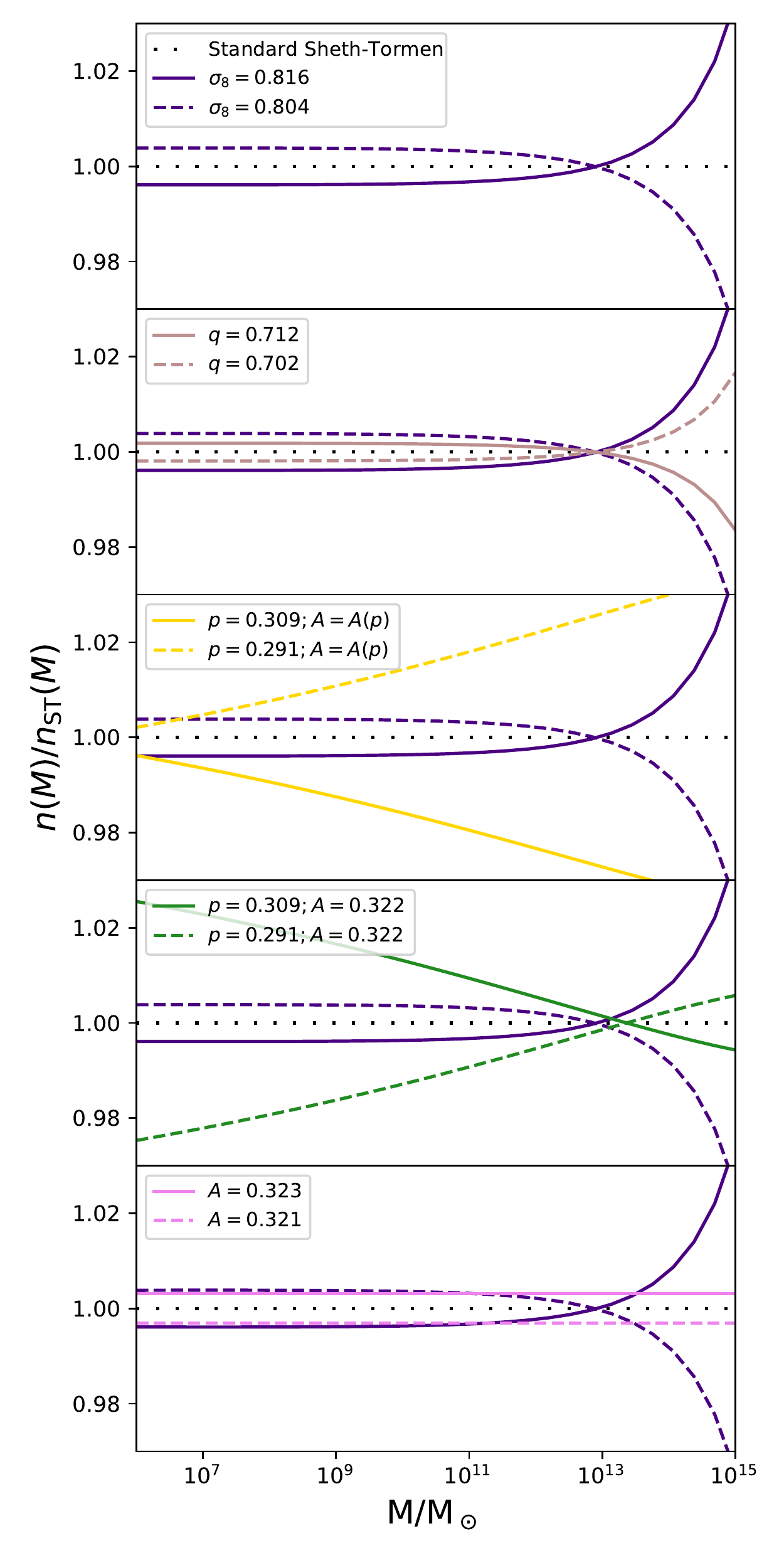}
\includegraphics[width=0.45\textwidth]{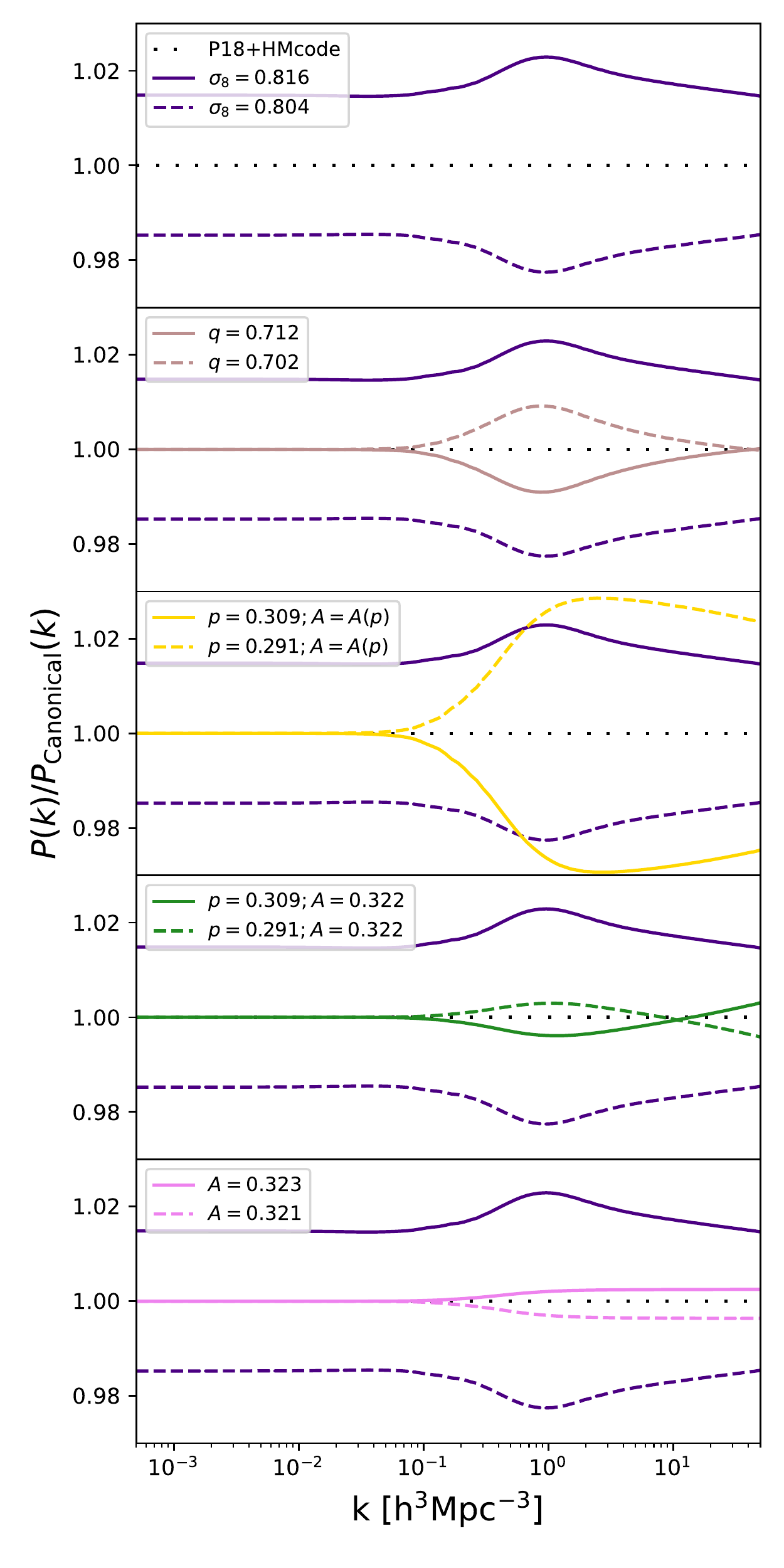}
\caption{Left Panels: The halo mass functions at $z = 0$ are shown with different HMF input parameter. The dotted line in the top-left sub-panel shows the canonical ST model with $q = 0.707$, $p = 0.3$, $\se = 0.81$, and $A(p) = 0.322$. The solid and dashed lines on this panel show the impact changing $\sigma_8$ (corresponding to a 1$\sigma$ uncertainty in \citet{planck18}), while keeping everything else unchanged. The other sub-panels on the left show the same lines as in the top-left sub-panel with two additional lines (a solid and a dashed) where some HMF parameters are changed. The modified values for $p$, $q$ and $A$ correspond to the uncertainty on these parameters from \citet{Despali+16}.\\
Right Panels: The \textsc{HMcode} 3-D matter power spectra at $z=0$ are shown with different HMF input parameters. The layout is the same as in the left panels, except that the right panels show the 3-D matter power spectra.}
\label{fig:RPk}
\end{figure*}

\label{secdt}
\begin{figure*}
\centering
    \includegraphics[width=0.45\textwidth]{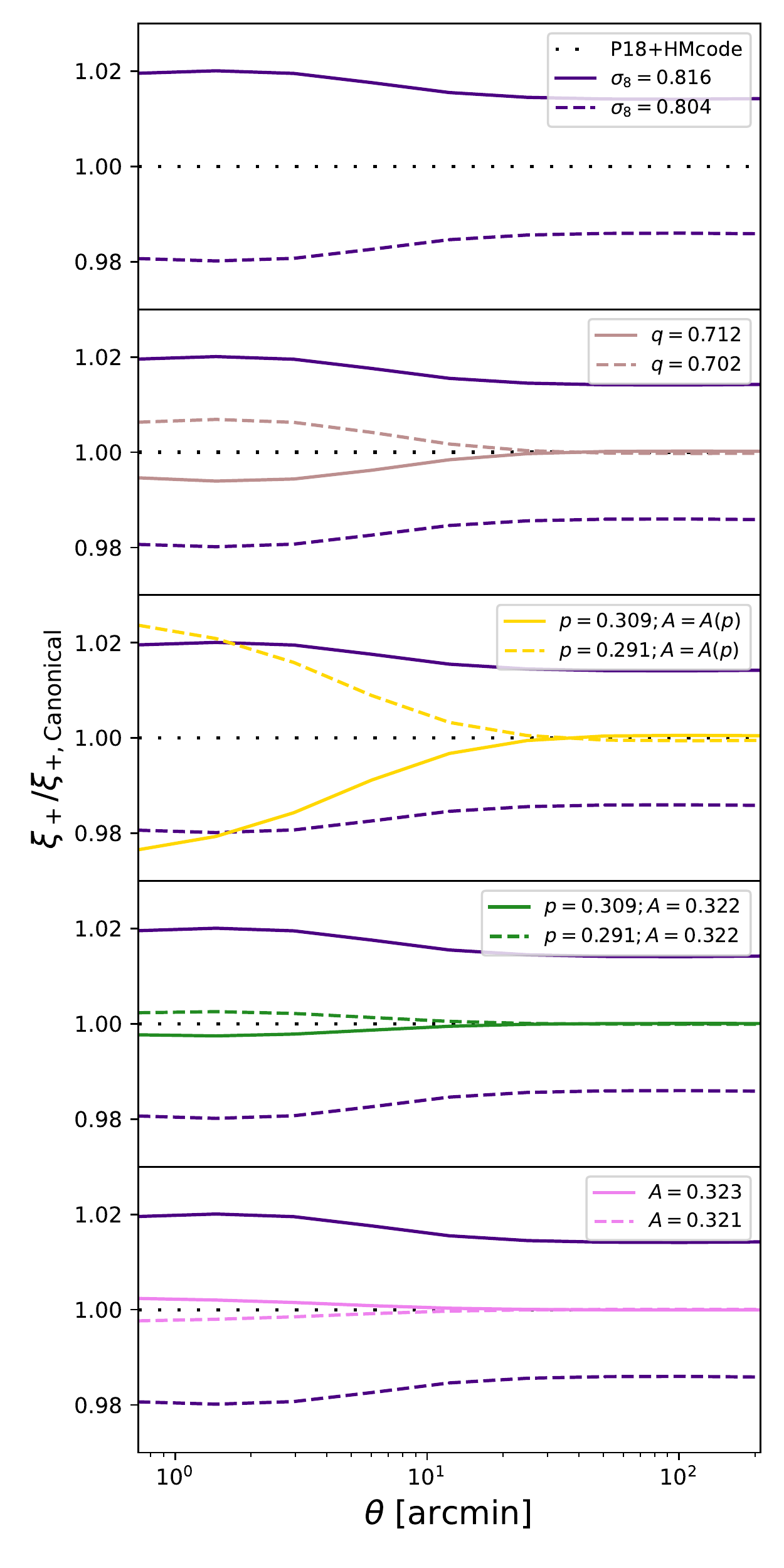}
    \includegraphics[width=0.45\textwidth]{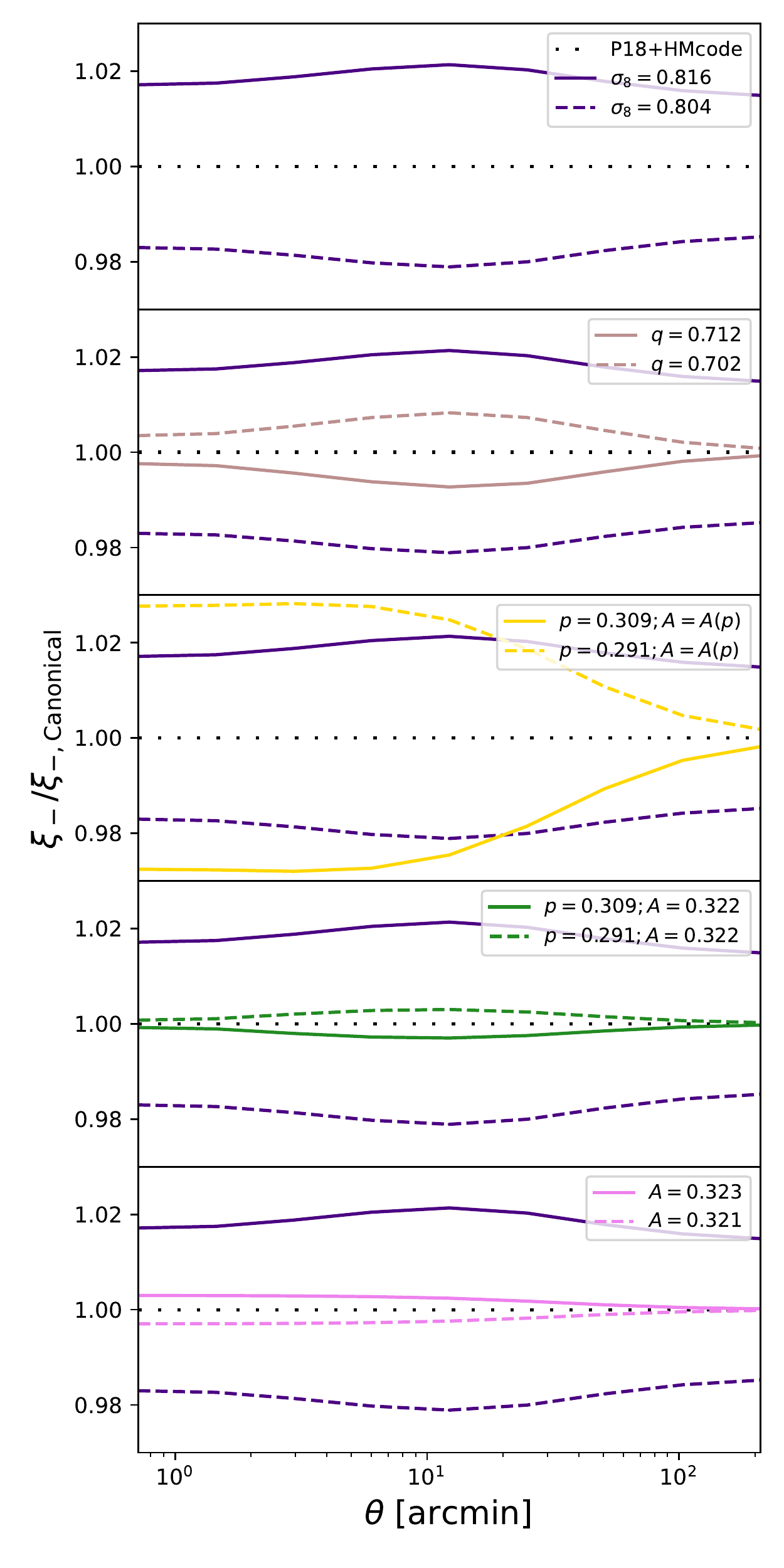}
\caption{Impact of modified HMF parameters on the shear correlation function $\xi_+$ (left panels) and $\xi_-$ (right panels). The layout and the curves is the same as in Figure \ref{fig:RPk}. The Limber projection is based on the redshift distribution of KiDS-1000 bin $5$.}
\label{fig:Rxi}
\end{figure*}

We use the \textit{Code for Anisotropies in the Microwave Background} (\textsc{CAMB}; \citealt{CAMB}) to compute the matter power spectrum $P_\delta$. The non-linear power spectra is calculated with our modified version of  \textsc{HMcode-2016} \citep{Mead+16} in order to include the HMF parameters $p$ and $q$ as free parameters.

The parameter space is sampled with the \textsc{multinest} algorithm, which is integrated in the \textsc{Cosmosis} framework \citep[][]{Zuntz+15}. \textsc{multinest} is a method that integrates over the posterior probability space, providing good precision and efficiency when dealing with high-dimensional parameter spaces. In order to characterize our chain outputs, we use the multinest-weighted mean instead of the maximum posterior, as it better depicts the Bayesian nature of the posterior distribution. It is worth noting that \textsc{multinest} may underestimate parameter errors, as shown by \citet{Lemos+22} where the error on $S_8$ can be underestimated by 5-10\%.

Each survey has its own specific redshift calibration, source selection, and corresponding scale cuts. For this reason, it is particularly interesting to compare how these two data sets will constrain the modified HMF. We use the \textit{\lcdm\ optimised} scale cut for DES-Y3. For KiDS-1000, we adopt the fiducial scale cut ($\theta(\xi_+) \in [0.5',300'], \theta(\xi_-) \in [4.0',300']$) (i.e. the one used by \citet{Asgari+21}). In Appendix \ref{asecsc}, we also provide additional KiDS-1000 scale cuts comparisons. To sample the HMF and profile parameters, we use the following flat priors:
\bea
2.0 \leq &B& \leq 3.13 \text{\ \citep[follows][]{Asgari+21}}, \\
-0.5 < &p& < 0.5, \\
0.0 < &q& < 2.0, \\
0.0 < &A& \leq 0.55, \text{\ unnormalised\ HMF\ only}.
\eea
For KiDS-1000, we use the same setup as \cite[][]{Asgari+21}, except for the \textsc{HMcode} and HMF parameters. For DES-y3, we made two modifications: we use \textsc{HMcode} and \textsc{Multinest} instead of \textsc{Halofit} and \textsc{Polychord} \citep[][]{Amon+22b,Secco+22}. 
In Figure \ref{fig:scale_cut_0} of Appendix \ref{asecsc}, we compare the DES-Y3 posterior contours between these two setups. The difference is small, but it is worth to mention, even if it is irrelevant to our discussion in this paper. \cite{DES+KiDS23} also found a similar difference between the two sampling codes. All other modelling choices, such as intrinsic alignments\footnote{In KiDS-1000 \citep{Asgari+21}, the intrinsic alignment model is the redshift-independent NLA model, while DES-y3 \citep{Amon+22a,Secco+22} uses the TATT model \citep{Blazek+19}.}, are kept unchanged compared to the original KiDS-1000 and DES-Y3 analyses.

Note that the amplitude $A$ of the HMF is occasionally considered as a free parameter, as initially proposed by \citet{Despali+16}. The question arises whether we should also set $A$ free in addition to ($p$, $q$). In the appendix of \citet{Mead+21}, it was demonstrated that abandoning the normalization requirement of the HMF does not affect the normalization of the one-halo term, implying that $A$ could be estimated if a sufficiently large range of scales is observed in addition to ($p$, $q$). We attempted to constrain ($p$, $q$, $A$) simultaneously, but discovered that $A$ and $q$ are strongly degenerate, and $p$ cannot be constrained. As a result, we employ A(p) from equation \ref{eqn:Ap} to normalize the HMF in our main analysis.

\section{Results}
\label{seccc}

\subsection{Impact of the halo mass function parameters on the halo abundance, $P(k)$ and shear correlation functions.}
\label{secpi}

\begin{table*}
\centering
\begin{tabular}{c|cccccc|c}
Cosmology & $\Om$ & $\Se$ & $h$ & $n_s$ & $M_\nu$ & $\Omega_b h^2$ & Reference \\
\hline
KiDS-1000 & 0.253 & 0.760 & 0.729 & 0.938 & 0.06 & 0.022 & \citet{Asgari+21} $\xi_\pm$ posterior mean \\
DES-y3 & 0.289 & 0.772 & 0.677 & 0.966 & 0.06 & 0.022 & \citet{Amon+22a,Secco+22} \lcdm-Optimised $\xi_\pm$ posterior mean \\
Planck 2018 & $0.316$ & $0.832$ & 0.677 & 0.966 & 0.06 & 0.022 & \citet{planck18} TT,TE,EE+lowE+lensing+BAO posterior mean \\
\hline
\end{tabular}
\caption{Cosmological parameters set-up of our study. All other independent cosmological parameters such as $N_{\nu,\mathrm{eff}}$ are fixed to the fiducial analysis and dependent cosmological parameters such $\Omega_\Lambda$ are varied with these parameters in order to keep a flat cosmology. In the case of DES, they did not report certain cosmological parameters, we used the {\it Planck} values for $n_s$ and $h$, since DES did not report what values they used.}
\label{tbl:cosmoparam}
\end{table*}

\begin{table*}
\centering
\begin{tabular}{c|ccc|c}
\# & $p$ & $q$ & $A$ & Note \bigstrut\\
\hline
ST1 & $0.300$ & $0.707$ & A(p) & \citet{STHMF} \bigstrut\\
ST2 & $0.360$ & $0.880$ & A(p) & A HMF parameter fit of the Model-ST1 HMF under the $200\rho_c$ Halo Mass Definition \bigstrut\\
ST3 & $0.402$ & $0.903$ & A(p) & Combination of ST4 and ST5 \bigstrut\\
ST4 & $0.402$ & $0.707$ & A(p) & Best \citet{planck18} $\Se$ recovery when fixing $q$ in KV450  \bigstrut\\
ST5 & $0.300$ & $0.903$ & A(p) & Best \citet{planck18} $\Se$ recovery when fixing $p$ in KV450 \bigstrut\\
ST6 & $0.000$ & $1.700$ & A(p) & Best \citet{planck18} $\Se$ recovery when fixing $p = 0$ in KV450 \bigstrut\\
ST7 & $0.237$ & $0.614$ & A(p) & A random-chosen high amplitude HMF  \bigstrut\\
\hline
DG1 & $0.258$ & $0.766$ & 0.330 & Original \citet{Despali+16} of virial haloes  \bigstrut\\
DG2 & $0.258$ & $0.766$ & 0.250 & Parameters giving a $25\%$ Lower HMF Amplitude \bigstrut\\
DG3 & $0.258$ & $0.766$ & 0.410 & Parameters giving a $25\%$ Higher HMF Amplitude  \bigstrut\\
\hline
ST-Free & Free & Free & A(p) & A flat HMF parameter prior based on the \citet{STHMF} Model \bigstrut\\
DG-Free & Free & Free & Free & A flat HMF parameter prior based on the \citet{Despali+16} Model \bigstrut\\
\hline
\end{tabular}
\caption{List of the various HMF models used in this paper. Models beginning with ST are based on the \citet{STHMF} halo mass function models, where A=A(p) indicates that the HMF is normalized using equation \ref{eqn:Ap}. Models beginning with DG are based on the \citet{Despali+16} HMF models, where the HMFs are not necessarily normalized. The reasons why we chose these particular HMF models are provided in the `Note' column of this table.}
\label{tbl:models}
\end{table*}

\begin{figure}
\centering
\includegraphics[width=0.49\textwidth]{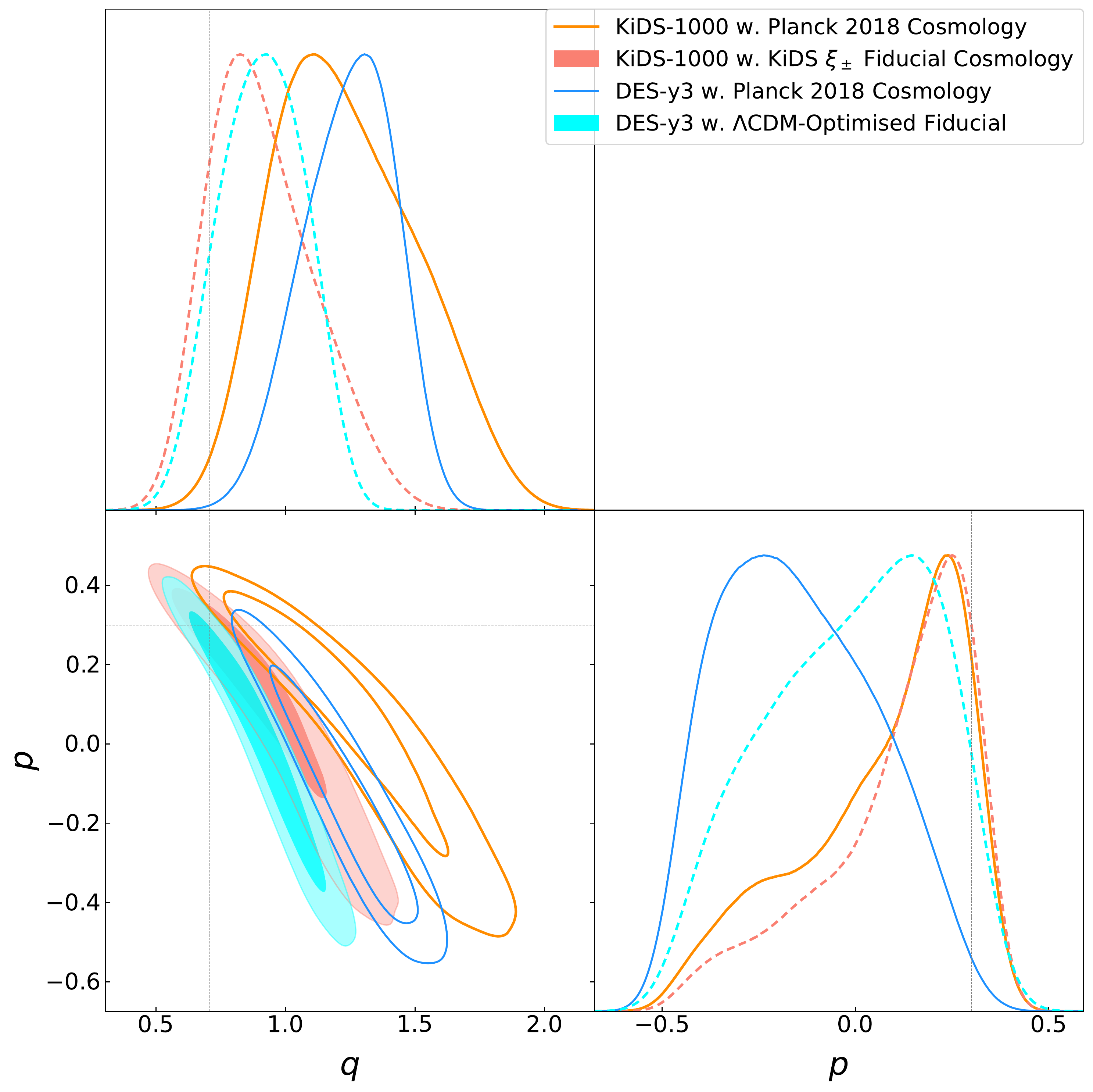}
\caption{Plot of the $q-p$ plane posterior contours for KiDS-1000 and DES-y3 data, with fixed cosmology either to their fiducial cosmic shear constraints (\salmon{salmon pink} for KiDS-1000 and \cyan{cyan} for DES-y3, shown as dashed line filled contours) or to the {\it Planck} 2018 (\darkorange{dark orange} for KiDS and \brightazure{bright azure} for DES, shown as solid line unfilled contours). All posteriors are sampled using \textsc{multinest}. The canonical ST parameters are indicated with a marker.}
\label{fig:pq_k1k}
\end{figure}

\begin{figure*}
\centering
\includegraphics[width=0.90\textwidth]{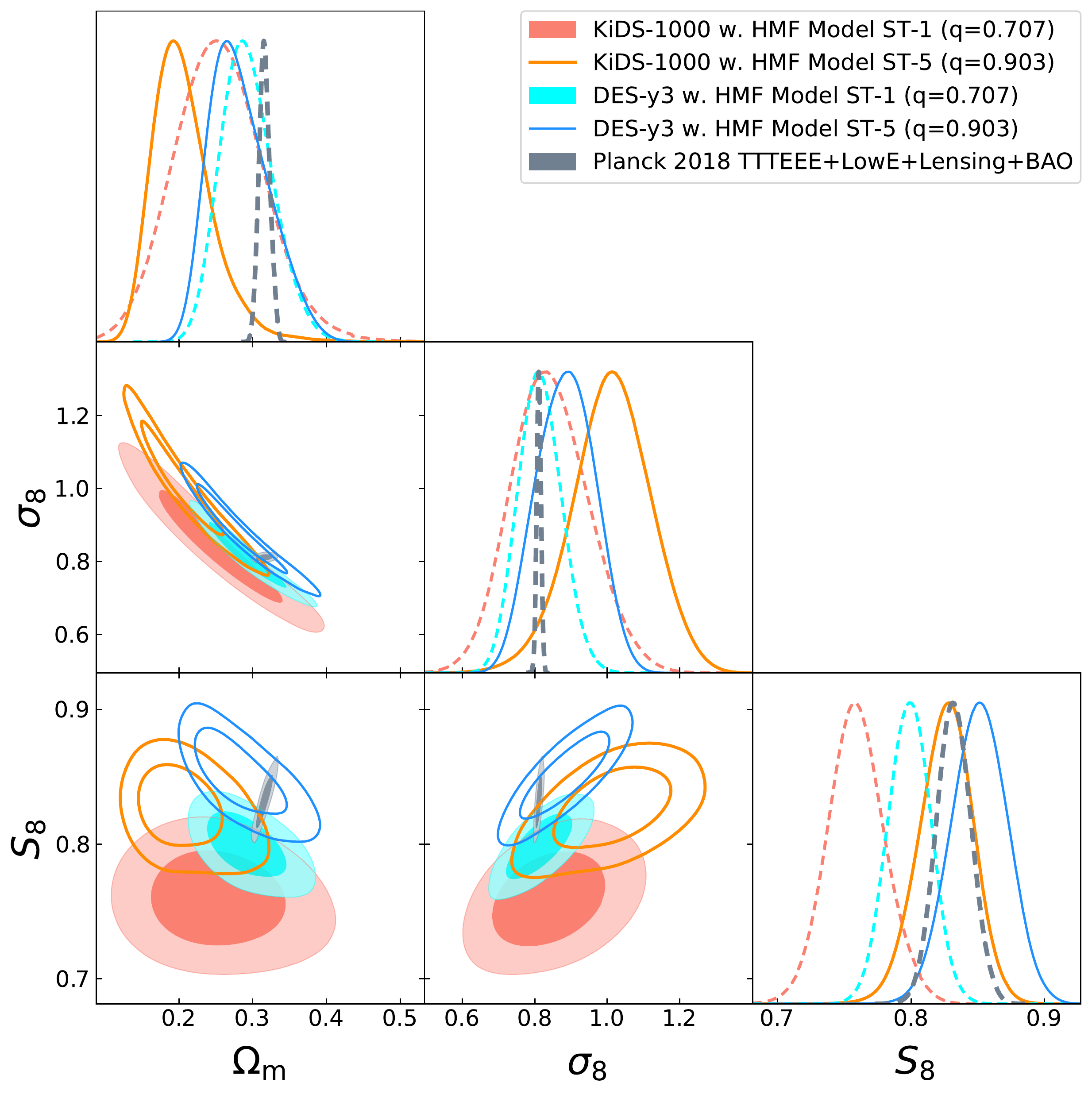}
\caption{Posterior contours ($68\%$ and $95\%$) in the $\Om-\se$, $\Om-\Se$ and $\se-\Se$ planes for KiDS-1000, DES-y3 and Planck. The \salmon{salmon-pink} filled contours (\citet{STHMF} halo mass function, labelled Model-\textit{ST1}) corresponds to the fiducial analysis of \citet{Asgari+21}. The solid \darkorange{dark orange} contour (Model-\textit{ST5}) correspond to $(p,q)=(0.3,0.903)$ (See table \ref{tbl:models}), the best \citet{planck18}-matching model in our table of halo mass function models. The filled \cyan{cyan} (DES-y3 with Model-\textit{ST1}) and solid \brightazure{bright azure} (DES-y3 with Model-\textit{ST5}) contours are posteriors obtained under the same logic from the DES-y3 pipeline with \textsc{HMcode} and \textsc{multinest}.
The \slategrey{slate grey} filled contours indicate the $95\%$ posterior of {\it Planck} 2018 TTTEEE+LowE+Lensing+BAO result \citep[][]{planck18} as the comparison.}\label{fig:cosmo_comp_k1k}
\end{figure*}

As discussed in Section \ref{HMFtheory}, the HMF in \citet{STHMF} is constructed using two parameters, $p$ and $q$, and an additional parameter $A$ if the normalisation is not fixed. In this section, we investigate the impact of a change of these parameters on the abundance of halos $n(M)$, the matter power spectrum $P(k)$ and the shear correlation functions $\xi_\pm$. This will allow us to visualise how the HMF degrees of freedom impact the quantities important for weak lensing calculations, and to what extent they might be degenerate with $\sigma_8$ \footnote{For weak lensing, $\Se$ would be the parameter to look at, but when $\Omega_m$ is fixed, $\sigma_8$ and $\Se$ are the same}. 

Figure \ref{fig:RPk} (left panels) shows the change in the HMF relative to the fiducial Sheth-Torman HMF (q = 0.707, p=0.3, A(p)=0.322) and Planck2018 cosmology when either $\sigma_8$ or Sheth-Torman parameters values are varied. The top-left sub-panel serves as a reference, where we show the matter power spectrum for three different values of $\sigma_8$ by $\pm 1\sigma$ using the \citet{planck18} as a fiducial value of 0.81 (horizontal dotted line). The corresponding dotted, solid and dashed curves are shown in all other left sub-panels for comparison to changes when the HMF parameters are varied. In these other left sub-panels, additional curves are shown with the HMF parameters varied within the uncertainty reported in \citet{Despali+16}. The second left sub-panel shows that varying $q$ by $\pm 0.05$ has a similar effect to $\sigma_8$, indicating that $q$ is partially degenerate with $\sigma_8$ as expected from Eqs. \ref{eqn:fnu} and \ref{haloterms}, since, in our formalism, only the 1-halo term depends on the HMF parameters. One can see that $q$ and $\sigma_8$ have a larger impact on the abundance of haloes with $M > 10^{13} M_\odot$. In the third left sub-panel, $p$ is varied by $0.009$ along with $A$ to keep the HMF normalised, while in the fourth left sub-panel, $A$ is fixed to its fiducial value while $p$ is varied by the same amount. Comparing these last two sub-panels, we observe that varying or fixing $A$ has a significant effect on the HMF. 

When $A$ is fixed, varying $p$ primarily impacts the abundance of low-mass haloes. When $A$ is used to normalise the HMF, haloes of all masses are impacted, with more sensitivity towards higher masses.
In this case a lower value of $p$ increases the abundance of haloes.  
Therefore, $p$ mostly impacts the abundance of halos with virial masses lower
than $10^{13}$ solar masses (when $A$ is fixed), while $q$ governs the abundance of
higher-mass-halos  ($M_\mathrm{Halo} > 10^{13} M_\odot$).

The right panels of Figure \ref{fig:RPk} show the variation in the \textsc{HMcode} matter power spectrum, using a similar plotting scheme as in the left panels. Changing the HMF parameters $p$ and/or $q$ results in a notable shape change of the matter power spectrum, particularly around the $k\sim1$ scale. Generally, in \textsc{HMcode}, the alteration of the halo models only affects the power spectrum at $k > 0.01$ because the two-halo term remains unaffected (see Eq. \ref{haloterms}). The second right sub-panel demonstrates that a higher $q$-value produces suppression of the power spectrum at the cluster scale ($k\sim1$) and a rise of power at small scales, resembling the baryonic feedback models \citep[][]{SchneiderTeyssier15}. It is interesting to note that $N$-body simulations favour a higher $q$-value compared to hydrodynamical simulations \citep[][]{Bocquet+16,Schaye+23}. On the other hand, increasing the parameter $p$ produces a power spectrum change up to $k = 10 \mathrm{h\ Mpc}^{-1}$ that more closely resembles a higher AGN feedback \citep[][]{Chisari+18}. This is due to the fact that the dark matter clustering at this scale is governed by the gas temperature of baryons \citep[][]{Beltz-MohrmannBerlind21}. Additionally, alternative dark matter models such as Axions have similar effects on this scale \citep[\eg,][]{Marsh15}.

Figure \ref{fig:Rxi} shows the shear correlation functions $\xi_\pm$ (equation \ref{eqn:shear_calc_transform}), following a similar layout as the panels in Figure \ref{fig:RPk} (The left panels of Figure \ref{fig:Rxi} are for $\xi_+$ and the right panels for $\xi_-$). It is observed that for angular scales $\theta < 30'$, changes in $p$ or $q$ have a similar effect as $\sigma_8$, but different scale dependence between $\xi_+$ and $\xi_-$. Here, all angular scales are affected, and changing $q$ has a more significant impact. This is the consequence of the fact that the shear correlation function is mixing all physical scales in the projection, where the clear separation between the 1-halo and the 2-halo terms in Fourier space is lost in the configuration space.
 
The comparison between the power spectra, the shear auto-correlation functions and the HMF (from figure \ref{fig:Rxi} and \ref{fig:RPk}), shows that modifying the HMF parameters leads to changes in these quantities which $\sigma_8$ cannot reproduce. This means that the HMF parameters and $\sigma_8$ are not degenerate, so we expect that the HMF parameters can be constrained in addition to $\sigma_8$. In the next section, we will explore quantitatively the cosmological parameters and the HMF parameters constraints with the DES and KiDS surveys.

We now present the results of our re-analysis of KiDS-1000 and DES-y3 data when both the halo mass function and cosmological parameters are permitted to vary either independently or jointly.

\begin{figure*}
\centering
\includegraphics[width=0.99\textwidth]{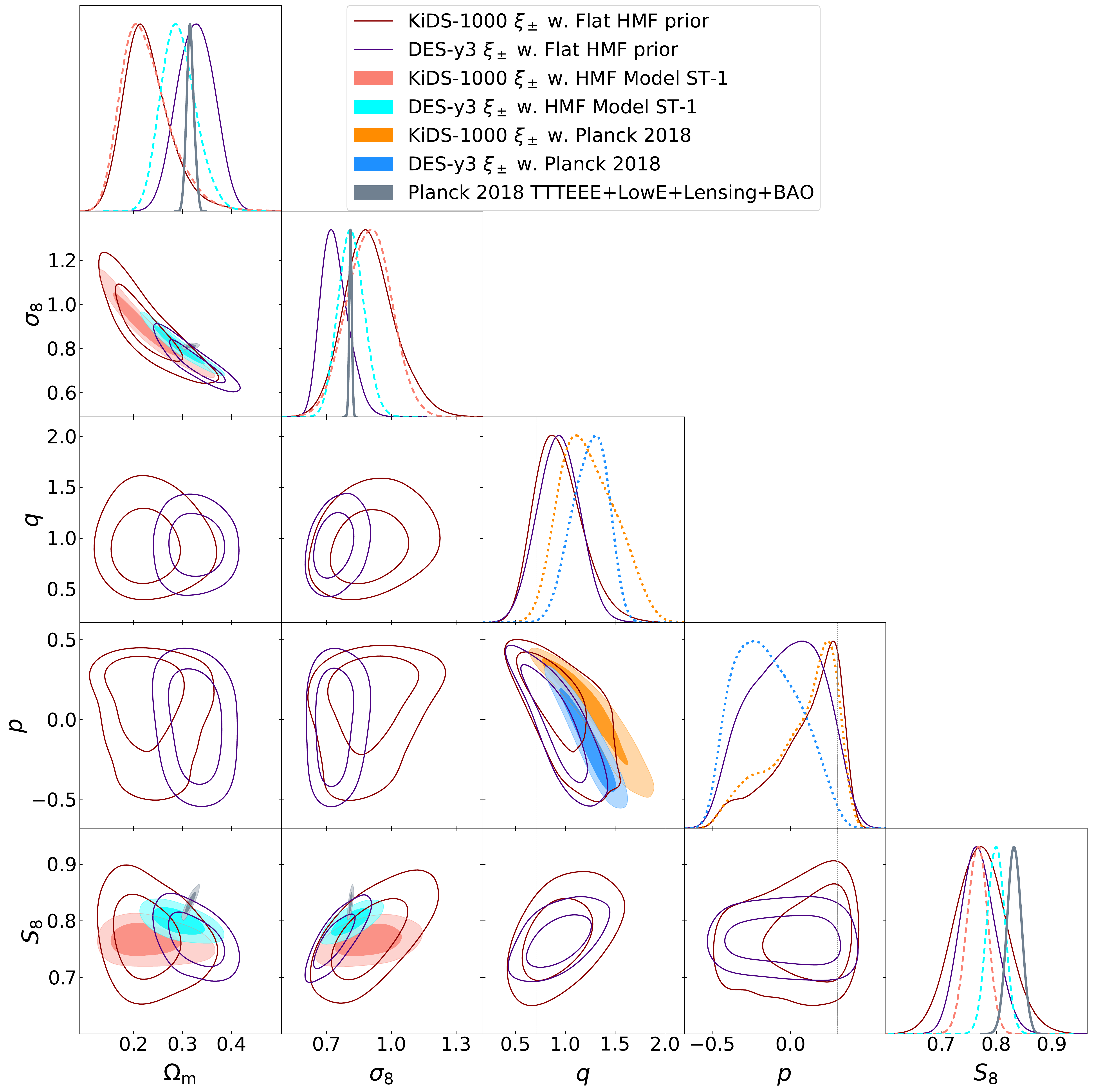}
\caption{Marginalised posterior contours for a flat $\Lambda$CDM model with the normalised \citet{STHMF} halo mass function model with the normalisation from the Equation \ref{eqn:Ap}. The solid \darkred{dark red} and \indigo{indigo blue} contours represent the results obtained by the KiDS-1000 and \textsc{HMcode}-based DES-y3 pipeline, respectively.
The \slategrey{slate grey} contours indicate the $95\%$ posterior of the {\it Planck} 2018 TTTEEE+LowE+Lensing+BAO result \citep[][]{planck18} for comparison.
We have also added the cosmological contours with fixed HMF Model-\textit{ST1} from Figure \ref{fig:cosmo_comp_k1k} and the HMF constraints with fixed {\it Planck} 2018 cosmology from Figure \ref{fig:pq_k1k} into this figure with their original colours as a comparison. The markers represent the fitted HMF parameters $p$ and $q$ from \citet{STHMF}, with the normalisation A(p) (equation \ref{eqn:Ap}).}
\label{fig:cosmo_ST99_degen}
\end{figure*}

\begin{table*}
\centering
\begin{tabular}{c|cccccc|cccc}
Model & KiDS $p$ & KiDS $q$ & KiDS $A$ & DES $p$ & DES $q$ & DES $A$ & KiDS $\Se$ & KiDS $\chi^2$ & DES $\Se$ & DES $\chi^2$  \bigstrut\\
\hline
ST1 & 0.300 & 0.707 & A(p) & 0.300 & 0.707 & A(p) & $0.765_{-0.020}^{+0.019}$ & 259.1 & $0.799_{-0.016}^{+0.016}$ & 282.7  \bigstrut\\
ST2 & 0.360 & 0.880 & A(p) & 0.360 & 0.880 & A(p) & $0.863_{-0.021}^{+0.024}$ & 257.8 & $0.867_{-0.023}^{+0.027}$ & 285.5   \bigstrut\\
ST3 & 0.402 & 0.903 & A(p) & 0.402 & 0.903 & A(p) & $0.920_{-0.031}^{+0.030}$ & 269.6 & $0.877_{-0.027}^{+0.027}$ & 288.2   \bigstrut\\
ST4 & 0.402 & 0.707 & A(p) & 0.402 & 0.707 & A(p) & $0.856_{-0.022}^{+0.024}$ & 257.6 & $0.852_{-0.025}^{+0.024}$ & 284.5   \bigstrut\\
ST5 & 0.300 & 0.903 & A(p) & 0.300 & 0.903 & A(p) & $0.826_{-0.019}^{+0.021}$ & 256.5 & $0.851_{-0.021}^{+0.020}$ & 284.0  \bigstrut\\
ST6 & 0.000 & 1.700 & A(p) & 0.000 & 1.700 & A(p) & $0.874_{-0.027}^{+0.027}$ & 273.2 & $0.887_{-0.022}^{+0.027}$ & 286.1  \bigstrut\\
ST7 & 0.257 & 0.614 & A(p) & 0.257 & 0.614 & A(p) & $0.702_{-0.018}^{+0.018}$ & 260.3 & $0.738_{-0.012}^{+0.011}$  & 282.4   \bigstrut\\
\hline
DG1 & 0.258 & 0.766 & 0.330 & 0.258 & 0.766 & 0.330 & $0.778_{-0.020}^{+0.019}$ & 258.9 & $0.812_{-0.017}^{+0.017}$  & 282.8   \bigstrut\\
DG2 & 0.258 & 0.766 & 0.250 & 0.258 & 0.766 & 0.250 & $0.823_{-0.021}^{+0.021}$ & 256.9 & $0.844_{-0.020}^{+0.020}$  & 284.0   \bigstrut\\
DG3 & 0.258 & 0.766 & 0.410 & 0.258 & 0.766 & 0.410 & $0.748_{-0.022}^{+0.022}$ & 264.4 & $0.783_{-0.015}^{+0.014}$ & 282.7   \bigstrut\\
\hline
Lensing & $0.11^{+0.24}_{-0.09}$ & $0.91^{+0.16}_{-0.24}$ & A(p) & $-0.01^{+0.28}_{-0.18}$ & $0.91^{+0.16}_{-0.17}$ & A(p) & 0.760 & 258.9 & 0.772 & 283.3   \bigstrut\\
 & $-0.03^{+0.28}_{-0.28}$ & $0.83^{+0.13}_{-0.18}$ & $0.34^{+0.07}_{-0.10}$ & $-0.01^{+0.26}_{-0.30}$ & $0.77^{+0.14}_{-0.14}$ & $0.35^{+0.10}_{-0.09}$ & 0.760 & 258.6 & 0.772 & 283.0   \bigstrut\\
\hline
Planck18 & $0.07^{+0.27}_{-0.12}$ & $1.23^{+0.24}_{-0.33}$ & A(p) & $-0.15^{+0.17}_{-0.26}$ & $1.25^{+0.20}_{-0.16}$ & A(p) & 0.832 & 263.4 & 0.832 & 284.8  \bigstrut\\
& $-0.02^{+0.27}_{-0.34}$ & $1.10^{+0.18}_{-0.23}$ & $0.35^{+0.08}_{-0.10}$ & $-0.04^{+0.21}_{-0.36}$ & $1.06^{+0.16}_{-0.16}$ & $0.42^{+0.11}_{-0.06}$ & 0.832 & 263.3 & 0.832 & 285.0  \bigstrut\\
\hline
ST-Free & $0.10^{+0.26}_{-0.10}$ & $0.93^{+0.19}_{-0.27}$ & A(p) & $-0.03^{+0.27}_{-0.22}$ & $0.93^{+0.21}_{-0.21}$ & A(p) & $0.771_{-0.046}^{+0.045}$ & 258.3 & $0.767_{-0.033}^{+0.029}$ & 282.2   \bigstrut\\
DG-Free & $0.01^{+0.27}_{-0.28}$ & $0.87^{+0.18}_{-0.22}$ & $0.37^{+0.08}_{-0.10}$ & $0.00_{-0.27}^{+0.25}$ & $0.76_{-0.17}^{+0.17}$ & $0.36^{+0.11}_{-0.01}$ & $0.782_{-0.053}^{+0.047}$ & 259.7 & $0.761_{-0.033}^{+0.031}$ & 282.6  \bigstrut\\
\hline
\end{tabular}
\caption{Table of posterior weighted-mean values of $\Se$ for each HMF model, along with the HMF parameters for each cosmological model, along with their corresponding marginal errors. The values are reported using weighted quantiles of $16\%$ and $84\%$ percentile of each parameter via the \textsc{Getdist} package \citep{getdist}. For the HMF models, A=A(p) indicates that the HMF is normalized via equation \ref{eqn:Ap}. Fixed parameters in the chain are reported without error bars. The $\chi^2$ value in the table corresponds to the minimum $\chi^2$ from the chain and may be strongly influenced by the randomness in the \textsc{multinest} sampler.}
\label{tbl:gofs}
\end{table*}

\subsection{Halo mass function constraints for fixed cosmology}
\label{ssecreshmf}

Here, we show the constraints on ($p$, $q$), for the fiducial and {\it Planck} 2018 cosmologies. For the KiDS and DES fiducial cosmologies, the mean posterior cosmology is assumed, which are $\xi_\pm$ posterior mean of \citet{Asgari+21} for KiDS-1000, Fiducial $\xi_\pm$ posterior mean from \citet{Amon+22a,Secco+22} for DES-y3 data. For the {\it Planck} 2018 cosmology, we use the TT,TE,EE+lowP+lensing+BAO posterior mean.
Although the cosmology is fixed for each data set, the nuisance parameters are still marginalized over, using the procedures described in \citet{Asgari+21}, \citet{Amon+22b}, and \citet{Secco+22}. Table \ref{tbl:cosmoparam} presents the cosmological parameters that were used.

The results are presented in Figure \ref{fig:pq_k1k}. Despite the degeneracy between $p$ and $q$, the standard values of the Sheth-Tormen (ST) model, denoted as Model-ST1, are still consistent with the lensing fiducial cosmologies of each survey. This is expected, it is shown here as a self-consistency test that the fiducial cosmology is not in tension with the ST1 values of $(p,q)$ even if they are allowed to vary.
However, when the {\it Planck} cosmology is assumed, the ST1 values are rejected at more than $2\sigma$. The {\it Planck} cosmology generally prefers a higher $q$ value. In the previous section, we showed that a higher $q$ corresponds to a lower abundance of high-mass haloes compared to ST1. One can show that, under the {\it Planck} cosmology, and for halo mass $M > 10^{14} M_\odot$, the posterior HMF has $29.5\%_{-8.5\%}^{+8.5\%}$ less mass in DES and $48.8\%_{-9.4\%}^{+8.3\%}$ less mass in KiDS compared to a standard \citet{STHMF} HMF.
The difference arises because the {\it Planck} cosmology predicts more high-mass haloes than what the weak lensing fiducial cosmologies suggest, a tension that could be addressed by changing the HMF parameters. While the amplitude of the matter power spectrum $\Se$ affects all scales of $P(k)$, the HMF parameters $p$ and $q$ only affect non-linear scales $k > 1$.
This result is consistent with the DES-y1 cluster abundance analysis \citep[][]{Costanzi+19b,DES+20Cluster}, which suggests that a richer mass abundance prediction requires a lower value of $\Se$ to compensate for the observation. One also notes that, while for the KiDS and DES-Y3 fiducial cosmologies the $(p,q)$ contours encompass the ST1 values, the same contours shift to higher $q$ and tend to split from each other with the {\it Planck} cosmology. It means that, in order to be consistent with the Planck18 cosmology, the best fit $P(k)$ model must be different between KiDS-1000 and DES-y3. This is of course an undesirable feature, and it is an illustration of how our approach can also be used as a self-consistency check between different surveys. It will be interesting to see how stage IV surveys (e.g. LSST versus Euclid) can perform on this test. In section \ref{sssecfhmf} we will show how precisely a LSST-type survey can measure $(p,q)$.

\begin{figure*}
\centering
\includegraphics[width=0.9\textwidth]{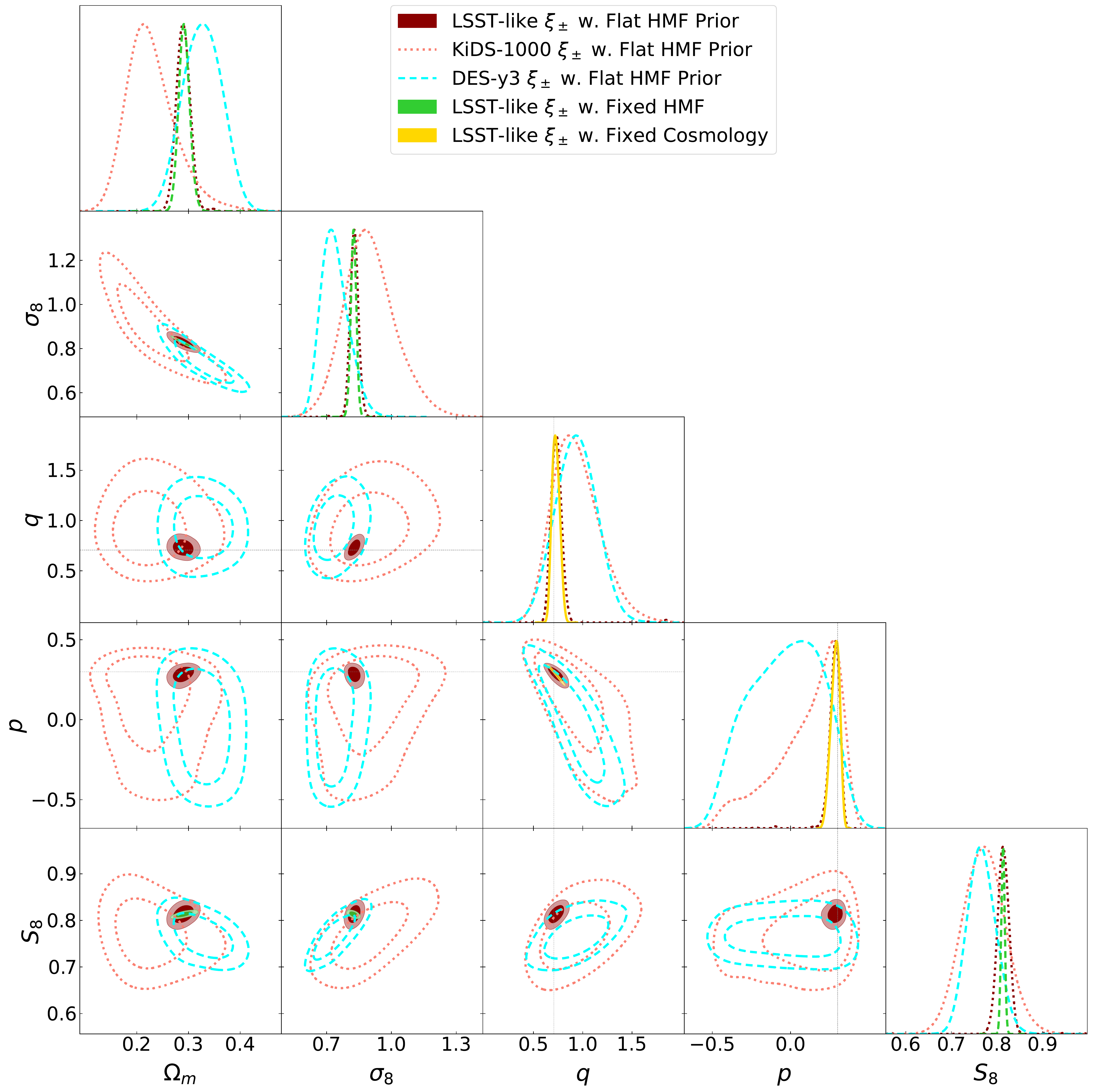}
\caption{The ($\Se$,$\Omega_m$) forecast for a LSST-like survey with corresponding Field-of-View and galaxy number density. The \darkred{darkred} contours correspond to a joint HMF and cosmology likelihood using an LSST-like setup. The \limegreen{limegreen} and \gold{gold} contours correspond to the LSST-like cosmological analysis with fixed $(p,q)=(0.3,0.707)$ and HMF-only analysis with a fixed cosmology the same as the SLICS simulation \citep[][]{Harnois-Deraps+18} respectively. These three sets of LSST-like contours use the \textsc{Pyccl}-generated theoretical mock data vectors and LSST-like covariance matrices computed from the SLICS simulation \citep[][]{Harnois-Deraps+18}. We also add both \salmon{KiDS-1000} and \cyan{DES-y3} joint HMF-cosmology contours here only in order to compare the constraining power between the Stage-III and the Stage-IV surveys.}
\label{fig:LSST}
\end{figure*}

\subsection{Cosmological constraints for fixed halo mass function parameters}

In order to test if a unique $n(M)$, common to KiDs and DES, is able to give consistent cosmological results (and whether the posterior cosmology might be consistent with Planck's or not), it would be useful to perform a series of tests with a set of fixed HMF parameters different from ST1.
The list of HMF parameters, labelled with numbers, that we will use are given in table \ref{tbl:models}. These models were determined during the exploration phase of our work using the KiDS-VIKING-450 data. Most of these models have higher $p$ and/or $q$ values than the original ST1 model. 
They were chosen so that $P(k,z=0)$ at $k \ge 1$ h Mpc$^{-1}$ is lowered by approximately $20\%$ to $30\%$ compared to the ST1 values, similar to some of the baryonic feedback models, such as gas outflows \citep[][]{Schneider+20}, counterbalancing the effect of a higher $\se$. This also corresponds to a deficit of roughly $30\%$ to $50\%$ of the high-mass end of the halo mass function ($M_\mathrm{Halo} > 10^{14} M_\odot$). The marginal errors on $(p,q)$ reported in Table \ref{tbl:gofs} correspond to $16\%$ and $84\%$ percentiles of each parameter calculated with the \textsc{Getdist} package \citep{getdist}.

Figure~\ref{fig:cosmo_comp_k1k} shows the contours for ST1 and ST5 (ST5 is chosen as it best matches the {\it Planck} $\Se$ from our list in Table \ref{tbl:models}). The solid blue and red contours show the cosmological constraints using ST1. The open contours show how the cosmological parameters posteriors are displaced with model ST5. As expected, with ST5, the change of HMF parameters lowers the abundance of high-mass haloes ($M_\mathrm{Halo} > 10^{13} M_\odot$) and therefore increases the inferred amplitude of $P(k)$].
Figure~\ref{fig:cosmo_comp_k1k} shows that looking at the shifts from the solid to the line contours, the $(p,q)$ degrees of freedom can easily lead to a few percent changes in the cosmological and HMF parameters. It is important to note that the new degrees of freedom do not make the model ill-constrained. This can be seen by looking at the $\chi^2$ values in Table \ref{tbl:gofs}, they do not dramatically change between the different models.
Models \textit{ST7} (see Table \ref{tbl:gofs}), which uses $(p,q)$ values lower than the ST1 model, also provides a good fit to both the KiDS-1000 and DES-y3 data, with a lower $\Se$, while the reduced $\chi^2$ remains very close to the of the original cosmological analysis with the HMF parameters fixed to ST1.
Table \ref{tbl:gofs} shows that, for the models we have chosen, the spectral features introduced by the different $(p,q)$ values are all acceptable by the data, which explains why, a change in $\Se$ can still absorb the changes caused by the different HMF parameters. We will see in the next subsection that this degeneracy is broken with stage IV surveys \footnote{Note that adding external, low redshift, cosmological prior from supernovae SNIa and Baryon Acoustic Oscillation would help in constraining $(p,q)$}. Our conclusion is that the KiDS-1000 and DES-y3 data have some spectral feature in the $P(k)$, essentially a power loss at the small scale, which a non-standard $(p,q)$ can absorb, but the constraining power of these surveys is not strong enough to reject low $(p,q)$ values. By comparing Figures \ref{fig:cosmo_comp_k1k} and \ref{fig:pq_k1k} we clearly understand why the Planck18 cosmology would definitely not work with the ST1 values: they are too low compared to the level of power loss that exists in the data.

\begin{figure*}
\centering
\includegraphics[width=0.45\textwidth]{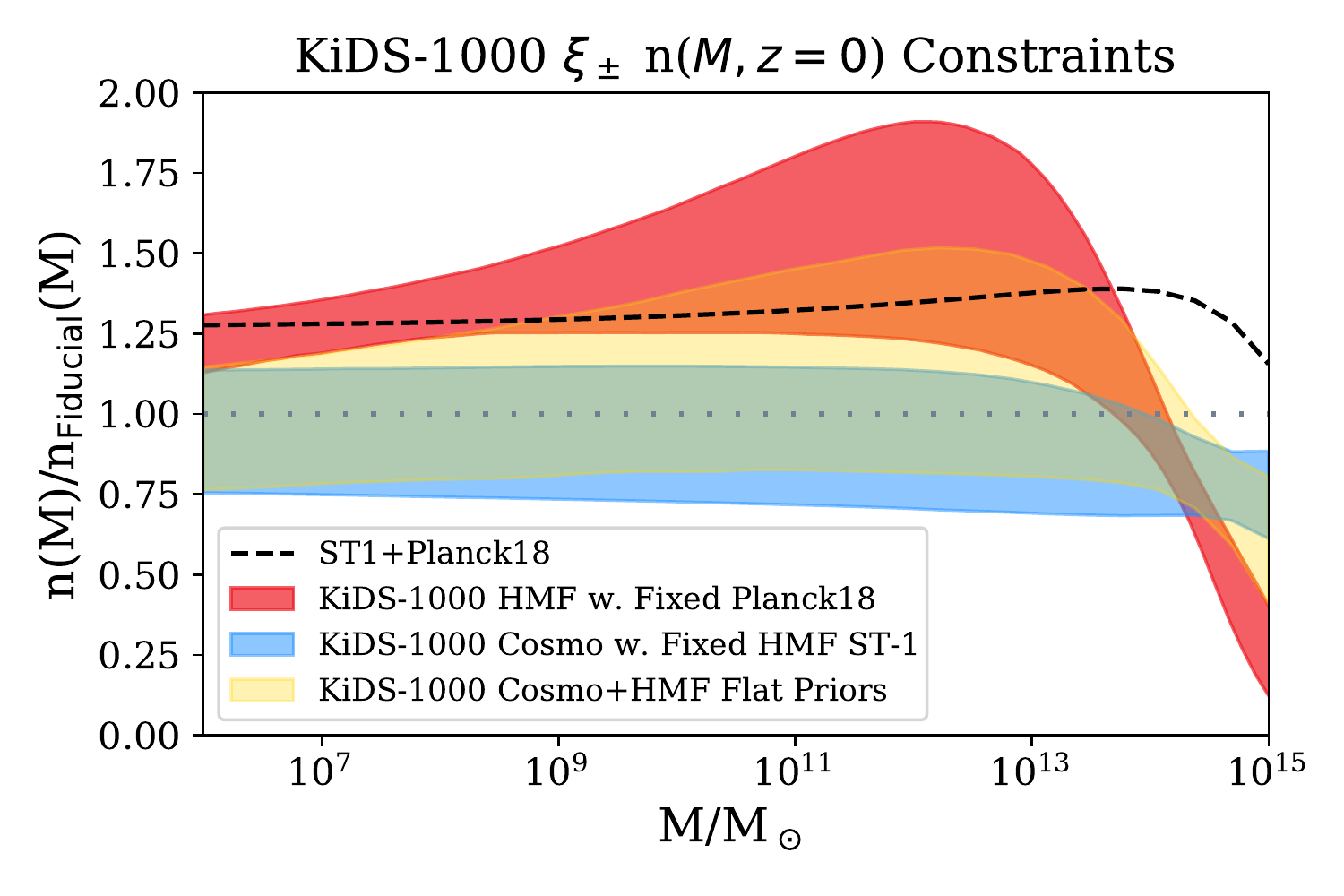}
\includegraphics[width=0.45\textwidth]{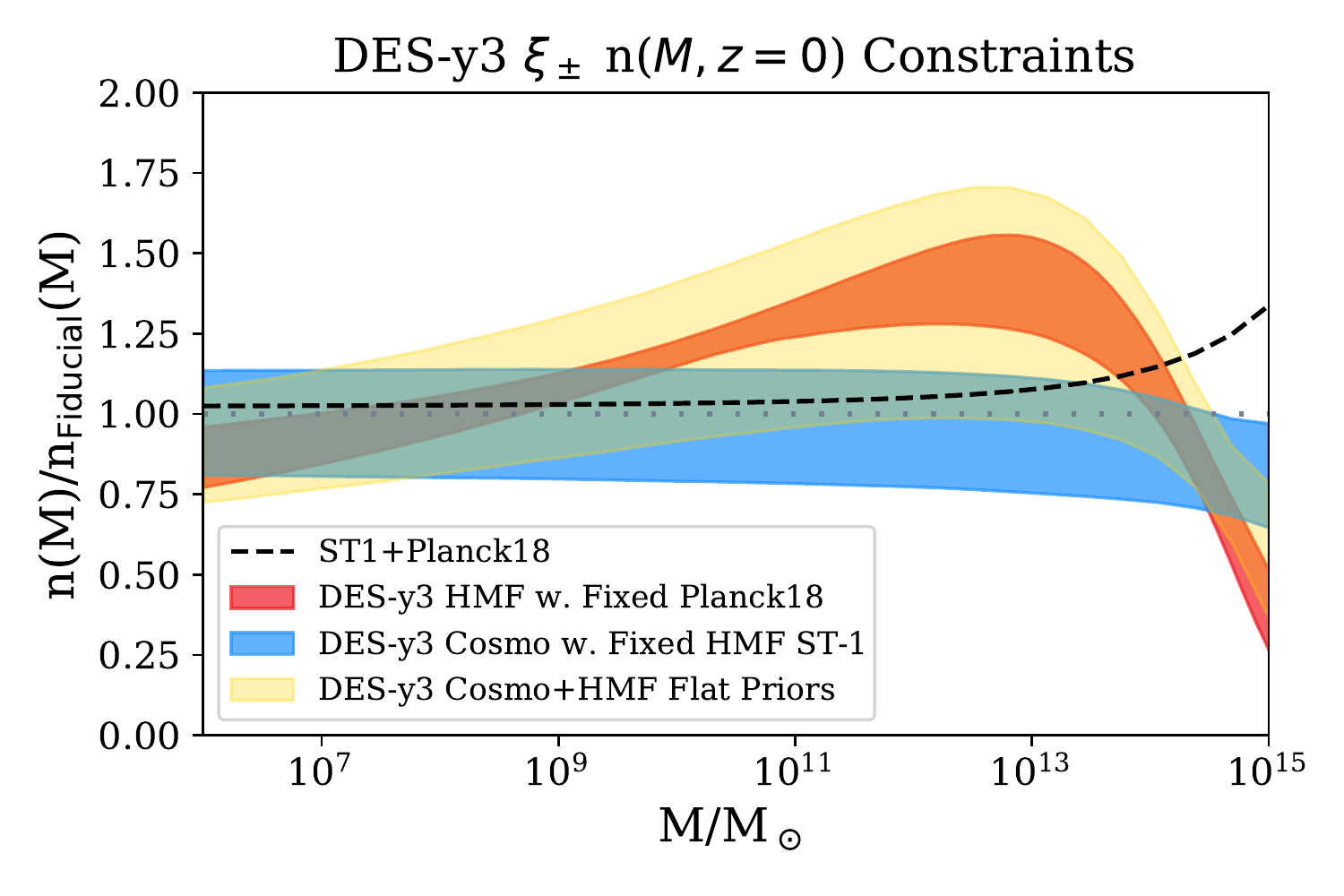}
\caption{Left panel: This panel shows the reconstructed halo mass function $n(M)$ obtained from the posteriors of this work using KiDS-1000 data, relative to the fiducial $n(M)$ calculated from the fiducial cosmological constraints of KiDS-1000 $\xi_\pm$ (labelled as `Fiducial'). The dashed line labelled as ST1+Planck18 corresponds to the Planck18 cosmology with standard Sheth-Tormen (ST) parameters $(p,q)=(0.3,0.707)$. The \chinared{red} contours correspond to the posterior obtained under a fixed Planck18 cosmology where only the ST parameters $(p,q)$ are allowed to vary, which corresponds to the \darkorange{dark orange} contour from figure \ref{fig:pq_k1k}. The \brightazure{bright azure} contour labelled as "KiDS-1000 Cosmo w. Fixed HMF ST1" corresponds to the fiducial cosmology chain where the cosmological parameters are allowed to vary but with a fixed Sheth-Tormen HMF, and this is represented by the \salmon{salmon pink} contour from figure \ref{fig:cosmo_comp_k1k}. Finally, the \gold{gold} contour labelled as "KiDS-1000 Cosmo+HMF Flat Priors" corresponds to the \darkred{dark red} contour of figure \ref{fig:cosmo_ST99_degen}, where both HMF and cosmological parameters are sampled under a flat prior. This plot is calculated for redshift $z=0$.
Right panel: The right panel shows the same information as the left panel, but obtained from the DES-y3 data.}
\label{fig:Nm_flatprior}
\end{figure*}

\begin{figure*}
\centering
\includegraphics[width=0.45\textwidth]{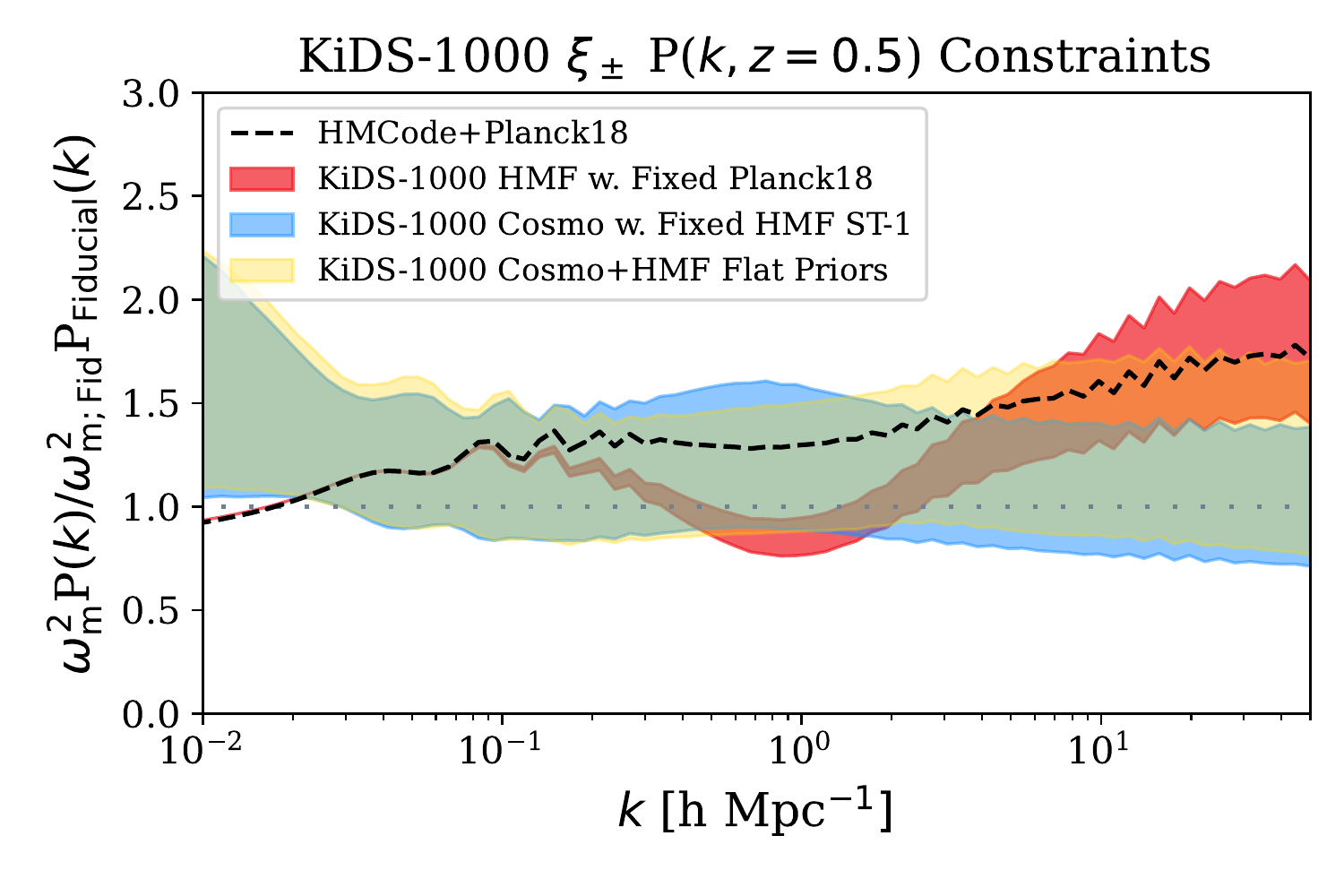}
\includegraphics[width=0.45\textwidth]{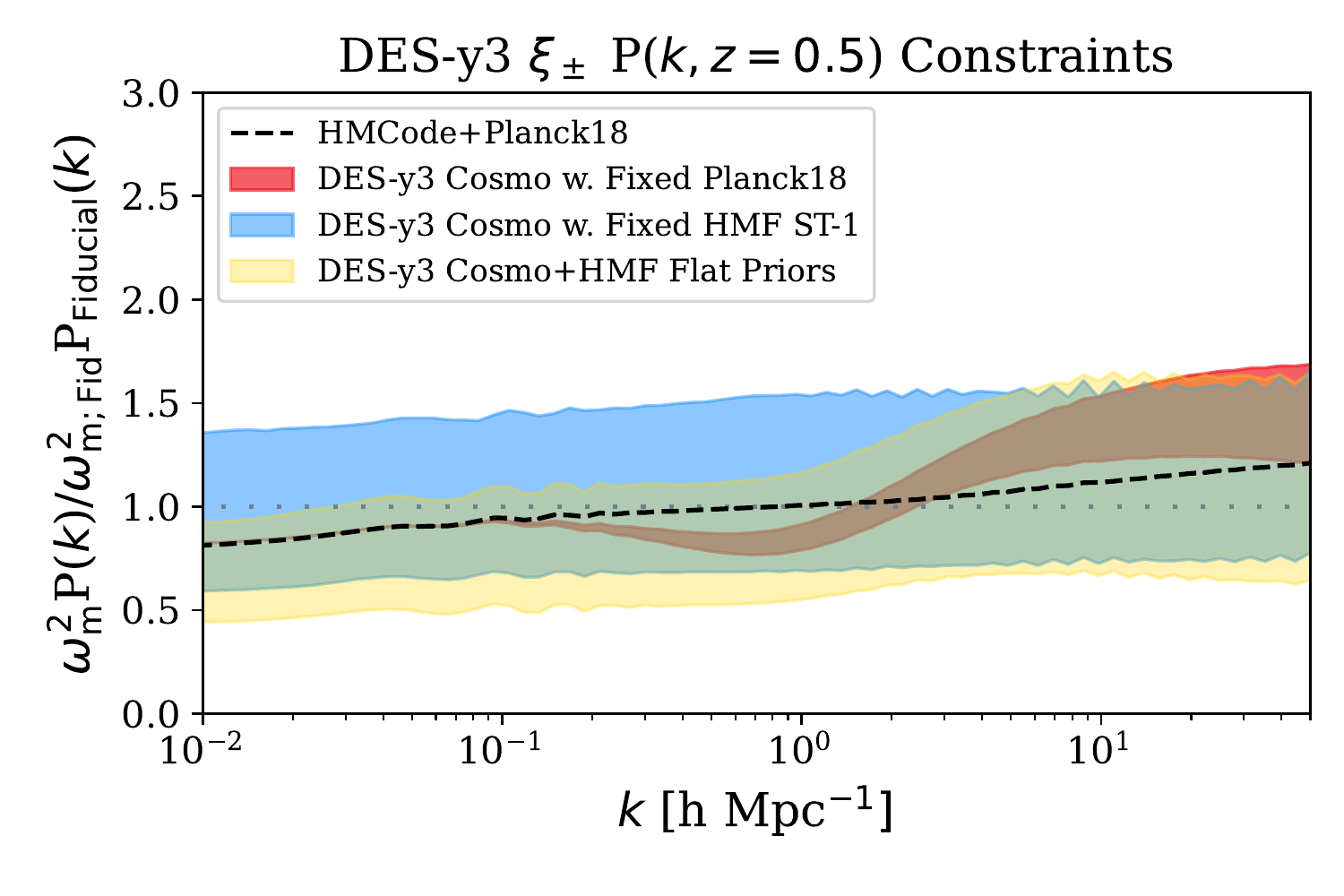}
\caption{The same as Fig \ref{fig:Nm_flatprior}, but for the reconstructed matter power spectrum $P(k)$ at redshift $z=0$ relative to the fiducial cosmologies and \textsc{HMcode} power spectra (labelled `Fiducial'). The normalisation factor of $\om = \Om h^2$ is used to represent the effect of lensing projections.} \label{fig:Pk_flatprior}
\end{figure*}

\subsection{Joint constraints on cosmological and Halo Mass Function parameters}
\label{sssecfhmf}

Having learned how $(p,q)$ impacts $P(k)$, and the interplay with the cosmological parameters, we can now explore the joint constraint on cosmology and the HMF.
Figure \ref{fig:cosmo_ST99_degen} shows the joint constraints on the HMF and cosmological parameters, with the flat priors. The $\Se$ posterior of the two surveys are in very good agreement, as well as the $(p,q)$ posteriors \footnote{Remember that it was not the case for the fixed Planck18 cosmology as the blue and orange contours in Figure \ref{fig:cosmo_ST99_degen} show}. The KiDS-1000 $\Se$ contour with flat HMF parameters prior is similar to Figure 5 in \cite{Amon+22c}, which shows that their $A_{\rm mod}$ parametrization and our modified HMF has a similar effect on the power spectrum. On the other hand, the DES-y3 with flat HMF prior tends to prefer a smaller $\sigma_8$ and larger $\Omega_m$ compared to their fiducial analysis, in a way which makes the $\Se$ posterior agrees with KiDS-1000. This is illustrated by the shift of $(\Omega_m,\sigma_8)$ contours which are both still aligned with the $\Se$ degeneracy. How can we interpret this result? It is known that moving along the $\Se$ degeneracy is particularly sensitive to the power spectrum slope. By comparing the $\chi^2$ values of the HMF flat prior cases to the fixed ST1 cases, for both KiDS-1000 and DES-y3 in Table \ref{tbl:gofs}, we clearly see that the $(p,q)$ degrees of freedom do not create a tension in the model, and, in fact, the $\chi^2$ are almost unchanged.

The meaning of Figure \ref{fig:cosmo_ST99_degen} is that with the HMF flat prior, the KiDS and DES contours moved such that $(p,q)$ and $\Se$ posteriors are in better agreement, but the distance between the $\Omega_m$ posteriors has increased. This behaviour could suggest an internal inconsistency in the model, revealed by the difficulty of finding a good agreement for the $(p,q), \Se$ and $\Omega_m$ posteriors simultaneously. The DES and KiDS contours on Figure \ref{fig:cosmo_ST99_degen} remain large and, at this stage, it only suggests a trend about how the contours want to move, rather than representing a definitive result.
We can however make the following comment: By introducing the HMF parameters, we have given the possibility for the best fit to deviate from the conventional theoretical wisdom $(p,q)=(0.3, 0.707)$. If this model is right, then the fit should recover the ST1 values within the errors, without creating tensions elsewhere. Figure \ref{fig:cosmo_ST99_degen} only suggests that it might not be the case, but more statistical power is necessary. The hybrid analysis in \cite{DES+KiDS23} leads to a reduction of the $\Se$ tension, without completely eliminating it. It would be interesting to rerun their analysis with our HMF flat prior setup and observe how the posteriors on $\Se$ and $\Omega_m$ want to move.

In order to see if future stage IV surveys could  measure the HMF parameters, we have performed a similar \textsc{Multinest}-based analysis on a Legacy Survey of Space and Time, LSST-like, survey. By using unbiased and noiseless mock data vectors generated by \textsc{Pyccl} \citep[][]{Chisari+19} for the fiducial cosmology of SLICS, we set constraints on parameters by analysing this mock data vector for fixed cosmological but variable HMF parameters, fixed HMF but variable cosmological parameters, and free HMF and cosmological parameters. 
We show the contours of these LSST-like contours, accompanied by current stage III KiDS-1000 and DES-y3 joint constraints on cosmological and HMF parameters, in Figure \ref{fig:LSST}.

Figure \ref{fig:LSST} shows that a stage IV survey can easily probe the HMF parameters in addition to probing cosmology. For instance, in the LSST-like set-up with flat prior of HMF parameters, the $\Se$ constraint reads $0.814_{-0.013}^{+0.012}$, while constraints of HMF parameters are $q = 0.741^{+0.049}_{-0.055}$ and $p =0.279^{+0.033}_{-0.032}$. Meanwhile, KiDS-1000 gives $\Se = 0.771_{-0.046}^{0.045}$, $q = 0.93^{+0.19}_{-0.27}$, and $p = 0.10^{+0.26}_{-0.10}$; DES-y3 also gives a very similar constraint to KiDS-1000 with $\Se = 0.767_{-0.033}^{+0.029}$, $q = 0.93^{+0.21}_{-0.21}$, and $p = -0.03^{+0.27}_{-0.22}$ -- stage III Errorbars on these parameters are roughly $\sim 4\times$ larger than our prediction of stage IV LSST-like surveys in both cosmological and HMF parameters. Note that we only used the Sheth and Tormen model, but our approach can easily be generalized to any halo model. We also notice that the widening of the LSST contours is significant when the HMF parameters are not fixed, \eg, the $S_8$ constraint widened from $S_8 = 0.813_{-0.004}^{+0.004}$, where the HMF parameters are fixed, to $S_8 = 0.814_{-0.013}^{+0.012}$, where the HMF parameters are not fixed. This is an important factor to consider for any kind of halo model being used for cosmological inference with stage IV surveys in the future.

\subsection{Full Posterior on the Halo Mass Function and $P(k)$}

In order to conclude our study, we have to quantify the range of changes in $n(M)$ and  $P(k)$
corresponding to the cosmology contours in Figure \ref{fig:cosmo_ST99_degen}.
This comparison is performed by taking the contours from Figures \ref{fig:pq_k1k}, \ref{fig:cosmo_comp_k1k} and \ref{fig:cosmo_ST99_degen} and calculating the corresponding $1-\sigma$ regions for $n(M)$ and $P(k)$. The results are shown on Figure \ref{fig:Nm_flatprior} for $n(M)$ and Figure \ref{fig:Pk_flatprior} for $P(k)$. Note that $P(k)$ is scaled by $\om = \Om h^2$ to account for projection effects. For each figure, the left panel is for KiDS-1000 and the right panel for DES-y3. In all panels, the horizontal line represents the normalization to the fiducial cosmology with the standard ST parameters. The black dashed line, on the other hand, shows the calculation for the Planck18 cosmology. The comparison between the dashed line and the horizontal dotted line gives another view of how the $\Se$ tension manifests itself in $n(M)$ and $P(k)$.
On all panels, the red shaded region shows the $1\sigma$ posterior of the ($p$, $q$) parameters with Planck18 cosmology from Figure \ref{fig:pq_k1k}. The bright azure region corresponds to the $1\sigma$ cosmological contours of both KiDS and DES under a canonical \citet{STHMF} (Figure \ref{fig:cosmo_comp_k1k}). Finally, the gold contours represent the corresponding $1\sigma$ region of Figure \ref{fig:cosmo_ST99_degen}, where both cosmological and HMF parameters are allowed to vary. All plots show calculations at redshift $z=0$.

Our analysis suggests that the DES $n(M)$ prefers a $29.5\%_{-8.5\%}^{+8.5\%}$ lower total mass and the KiDS $n(M)$ prefers a $48.8\%_{-9.4\%}^{+8.3\%}$ lower total mass for all halos with masses $M_\mathrm{Halo} > 10^{14} M_\odot$, compared to \citet{planck18} cosmological model with the standard \citet{STHMF} HMF parameters. This agrees with recent studies on halo mass abundance (\eg, \citealt{Li+19, DES+20Cluster}), which both suggest astrophysical explanations for this low abundance. 
Regarding $P(k)$, the red region in Figure \ref{fig:Pk_flatprior} shows a power spectrum suppression by $\sim 25\%$ at $k \simeq 1 \mathrm{h\ Mpc}^{-1}$ for the Planck18 cosmology. This is a very strong amplitude suppression because it comes in addition to the internal modifications of the halo model in 
\textsc{HMcode}, which accounts for the baryonic effect. Interestingly, our result is similar to the value $A_{\rm mod}\sim 0.7$ obtained from Equation \ref{Amon} found by \cite{Amon+22c}. In other words, the unrealistic high Active Galactic Nuclei (AGN) temperature reported by \cite{Amon+22c} is also found with the non-AGN version of \textsc{HMcode} when the HMF parameters are free to vary.
Galaxy-galaxy lensing (GGL) encounters a similar issue, known as the 'Lensing-is-Low' problem. \cite[][]{Leauthaud+17} found that the measured surface mass density contrast $\Delta\Sigma_\mathrm{data}$ at small scale is approximately 20\% smaller than the theoretical predictions if we assume the Planck18 cosmology. In comparison, for KiDS-1000, \citep[][]{Amon+22b} showed that, assuming the fiducial lensing cosmology from \citep[][]{Heymans+21}, the data--to-theory ratio $\Delta\Sigma_\mathrm{data}/\Delta\Sigma_\mathrm{theory}$ is close to one. This result, again, outlines a fundamental incompatibility between the Planck18 cosmology and the shape of $P(k)$.

\section{Conclusion}
\label{seccl}

In this study, we used weak gravitational lensing data from KiDS-1000 and DES-Y3 to constrain the cosmological parameters and the HMF. The $(p,q)$ parameters add two new degrees of freedom to the calculation of the matter power spectrum $P(k)$. 
We examined various HMF configurations, where the cosmological parameters were either fixed to specific values (KiDS, DES, or {\it Planck} cosmologies) or allowed to vary. To maintain the maximum consistency of our analysis with the original KiDS-1000 and DES-Y3 studies, we used their original pipelines and only made minimal changes to the samplers and modelling choices, as outlined in section \ref{secdt}.

We modified \textsc{HMcode} \citep{Mead+15, Mead+16} to account for these changes. We first assessed how different parameter choices impact the Sheth-Tormen HMF \citep{STHMF}, the matter power spectrum $P(k)$ and the shear correlation function $\xi_\pm$. We found that the amplitude $\Se$ is generally higher when both the cosmology and $(p,q)$ are free. The new $\Se$ posterior is in 
better agreement with the Planck18 value, and we do recover the ST1 $(p,q)$ when allowing for both cosmological and HMF parameters to vary, but at the expense of reducing consistency along the $(\sigma_8,\Omega_m)$ degeneracy. Moreover, $P(k)$ must be reduced by 25\% at scale $k\sim 1~{\rm h/Mpc}$, in agreement with other lensing studies. 
We should be cautious about the interpretation of this result: this is not a measurement of the mass function, nor a proof that it should be modified: if we found enough evidence that p and q are not equal to ST1 values, then that would indicate a missing component in the modelling, but since the modelling is tuned to the simulations, it also indicates a missing factor in the simulations, which may or may not be related to HMF.
This is what is happening for the Planck18 cosmology, where the $(p,q)$ contours strongly exclude the ST1 values. A useful analogy is the shear calibration parameter $m$, which should be zero if no residual systematics are present in the shear measurement, but all lensing surveys generally find a small but non-zero value, indicating a bias in the raw shear measurement.

It is interesting that \cite{DES+KiDS23} found that their hybrid pipeline reconciles DES-y3, KiDS-1000 and Planck18 by a unified set of choices for the intrinsic alignment, baryonic feedback, the choice of angular scales and halo model code. It would be very interesting to see if, with this setup, the $(p,q)$ parameters would converge to $(0.3,0.707)$, and this is left for future work.

The $\Se$ tension has triggered a lot of excitement about new physics because the residual systematics coming from shape measurement and photometric redshift calibration seem to be well under control. 
Unfortunately, we have to admit that the understanding of the theory (and specifically the calculation of $P(k)$ and $C_\ell$ with all the complexities) is now lagging behind the constraining power of existing lensing surveys.
This will be even worse for stage IV surveys, therefore there is an urge to make progress in this area. Fortunately, the potential of weak lensing science is still vastly under-utilized on data: there is a lot of constraining power on intrinsic alignment, baryonic physics and the non-linear growth of structures which can come from combining high order statistics, peak statistics, cross-correlations, and what we could call "global statistics" such as void- lensing and density split statistics. Almost no effort has been done so far to use these tools together to constrain the complexities of $P(k)$, this is the next frontier.

\section*{Acknowledgements}

We thank Gary Hinshaw and Robert Reischke for fruitful discussions. SG and LVW acknowledge support by the University of British Columbia, Canada's NSERC. TT acknowledges funding from the Leverhulme Trust and the Swiss National Science Foundation under the Ambizione project PZ00P2\_193352. 
 
\textit{Author contributions:}
All authors have contributed to the development and writing of this paper. The authorship list is divided into three groups: the lead authors (SG, MAD, \& LW) followed by two alphabetical groups. The first alphabetical group (MA, AM, TT) includes key contributors to both the scientific analysis and the data products. The second group (ZY) includes those who have either made significant contributions to the data products or the scientific analysis.

\section*{Data Availability}
 
In this paper the data is analysed with these open-source python packages:
\textsc{astropy} \citep{astropy13,astropy18},
\textsc{Camb} \citep{CAMB},
\textsc{Cosmosis} \citep{Zuntz+15},
\textsc{Getdist} \citep{getdist},
\textsc{KCAP} \citep{Troester+21},
\textsc{matplotlib} \citep{matplotlib},
\textsc{Multinest} \citep{Multinest}, 
\textsc{numpy} \citep{numpy}, 
\textsc{pyccl} \citep{Chisari+19},
\textsc{pyHMcode} \citep{Troester+22},
and \textsc{scipy} \citep{scipy}. The data analysis pipeline and generated posterior chain files used in this article will be shared on reasonable request
to the corresponding author (SG).



\bibliographystyle{mnras}
\bibliography{reference} 

\begin{thebibliography}{}
\makeatletter
\relax
\def\mn@urlcharsother{\let\do\@makeother \do\$\do\&\do\#\do\^\do\_\do\%\do\~}
\def\mn@doi{\begingroup\mn@urlcharsother \@ifnextchar [ {\mn@doi@}
  {\mn@doi@[]}}
\def\mn@doi@[#1]#2{\def\@tempa{#1}\ifx\@tempa\@empty \href
  {http://dx.doi.org/#2} {doi:#2}\else \href {http://dx.doi.org/#2} {#1}\fi
  \endgroup}
\def\mn@eprint#1#2{\mn@eprint@#1:#2::\@nil}
\def\mn@eprint@arXiv#1{\href {http://arxiv.org/abs/#1} {{\tt arXiv:#1}}}
\def\mn@eprint@dblp#1{\href {http://dblp.uni-trier.de/rec/bibtex/#1.xml}
  {dblp:#1}}
\def\mn@eprint@#1:#2:#3:#4\@nil{\def\@tempa {#1}\def\@tempb {#2}\def\@tempc
  {#3}\ifx \@tempc \@empty \let \@tempc \@tempb \let \@tempb \@tempa \fi \ifx
  \@tempb \@empty \def\@tempb {arXiv}\fi \@ifundefined
  {mn@eprint@\@tempb}{\@tempb:\@tempc}{\expandafter \expandafter \csname
  mn@eprint@\@tempb\endcsname \expandafter{\@tempc}}}

\bibitem[\protect\citeauthoryear{{Abbott} et~al.,}{{Abbott}
  et~al.}{2020}]{DES+20Cluster}
{Abbott} T.~M.~C.,  et~al., 2020, \mn@doi [\prd] {10.1103/PhysRevD.102.023509},
  \href {https://ui.adsabs.harvard.edu/abs/2020PhRvD.102b3509A} {102, 023509}

\bibitem[\protect\citeauthoryear{{Aihara} et~al.,}{{Aihara}
  et~al.}{2018}]{Aihara+18}
{Aihara} H.,  et~al., 2018, \mn@doi [\pasj] {10.1093/pasj/psx066}, \href
  {https://ui.adsabs.harvard.edu/abs/2018PASJ...70S...4A} {70, S4}

\bibitem[\protect\citeauthoryear{{Aiola} et~al.,}{{Aiola}
  et~al.}{2020}]{Aiola+20}
{Aiola} S.,  et~al., 2020, arXiv e-prints, \href
  {https://ui.adsabs.harvard.edu/abs/2020arXiv200707288A} {p. arXiv:2007.07288}

\bibitem[\protect\citeauthoryear{{Amon} \& {Efstathiou}}{{Amon} \&
  {Efstathiou}}{2022}]{Amon+22c}
{Amon} A.,  {Efstathiou} G.,  2022, arXiv e-prints, \href
  {https://ui.adsabs.harvard.edu/abs/2022arXiv220611794A} {p. arXiv:2206.11794}

\bibitem[\protect\citeauthoryear{{Amon} et~al.,}{{Amon}
  et~al.}{2022a}]{Amon+22b}
{Amon} A.,  et~al., 2022a, arXiv e-prints, \href
  {https://ui.adsabs.harvard.edu/abs/2022arXiv220207440A} {p. arXiv:2202.07440}

\bibitem[\protect\citeauthoryear{{Amon} et~al.,}{{Amon}
  et~al.}{2022b}]{Amon+22a}
{Amon} A.,  et~al., 2022b, \mn@doi [\prd] {10.1103/PhysRevD.105.023514}, \href
  {https://ui.adsabs.harvard.edu/abs/2022PhRvD.105b3514A} {105, 023514}

\bibitem[\protect\citeauthoryear{{Asgari} et~al.,}{{Asgari}
  et~al.}{2021}]{Asgari+21}
{Asgari} M.,  et~al., 2021, \mn@doi [\aap] {10.1051/0004-6361/202039070}, \href
  {https://ui.adsabs.harvard.edu/abs/2021A&A...645A.104A} {645, A104}

\bibitem[\protect\citeauthoryear{{Asgari}, {Mead}  \& {Heymans}}{{Asgari}
  et~al.}{2023}]{Asgari+23}
{Asgari} M.,  {Mead} A.~J.,   {Heymans} C.,  2023, \mn@doi [arXiv e-prints]
  {10.48550/arXiv.2303.08752}, \href
  {https://ui.adsabs.harvard.edu/abs/2023arXiv230308752A} {p. arXiv:2303.08752}

\bibitem[\protect\citeauthoryear{{Astropy Collaboration} et~al.,}{{Astropy
  Collaboration} et~al.}{2013}]{astropy13}
{Astropy Collaboration} et~al., 2013, \mn@doi [\aap]
  {10.1051/0004-6361/201322068}, \href
  {https://ui.adsabs.harvard.edu/abs/2013A&A...558A..33A} {558, A33}

\bibitem[\protect\citeauthoryear{{Astropy Collaboration} et~al.,}{{Astropy
  Collaboration} et~al.}{2018}]{astropy18}
{Astropy Collaboration} et~al., 2018, \mn@doi [\aj] {10.3847/1538-3881/aabc4f},
  \href {https://ui.adsabs.harvard.edu/abs/2018AJ....156..123A} {156, 123}

\bibitem[\protect\citeauthoryear{{Bagla}, {Khandai}  \& {Kulkarni}}{{Bagla}
  et~al.}{2009}]{Bagla+09}
{Bagla} J.~S.,  {Khandai} N.,   {Kulkarni} G.,  2009, arXiv e-prints, \href
  {https://ui.adsabs.harvard.edu/abs/2009arXiv0908.2702B} {p. arXiv:0908.2702}

\bibitem[\protect\citeauthoryear{{Bahcall} \& {Cen}}{{Bahcall} \&
  {Cen}}{1993}]{BahcallCen93}
{Bahcall} N.~A.,  {Cen} R.,  1993, \mn@doi [\apjl] {10.1086/186803}, \href
  {https://ui.adsabs.harvard.edu/abs/1993ApJ...407L..49B} {407, L49}

\bibitem[\protect\citeauthoryear{{Barreira}, {Li}, {Hellwing}, {Lombriser},
  {Baugh}  \& {Pascoli}}{{Barreira} et~al.}{2014}]{Barreira+14}
{Barreira} A.,  {Li} B.,  {Hellwing} W.~A.,  {Lombriser} L.,  {Baugh} C.~M.,
  {Pascoli} S.,  2014, \mn@doi [\jcap] {10.1088/1475-7516/2014/04/029}, \href
  {https://ui.adsabs.harvard.edu/abs/2014JCAP...04..029B} {2014, 029}

\bibitem[\protect\citeauthoryear{{Baugh} et~al.,}{{Baugh}
  et~al.}{2019}]{Baugh+19}
{Baugh} C.~M.,  et~al., 2019, \mn@doi [\mnras] {10.1093/mnras/sty3427}, \href
  {https://ui.adsabs.harvard.edu/abs/2019MNRAS.483.4922B} {483, 4922}

\bibitem[\protect\citeauthoryear{{Beltz-Mohrmann} \&
  {Berlind}}{{Beltz-Mohrmann} \& {Berlind}}{2021}]{Beltz-MohrmannBerlind21}
{Beltz-Mohrmann} G.~D.,  {Berlind} A.~A.,  2021, arXiv e-prints, \href
  {https://ui.adsabs.harvard.edu/abs/2021arXiv210305076B} {p. arXiv:2103.05076}

\bibitem[\protect\citeauthoryear{{Ben{\'\i}tez}}{{Ben{\'\i}tez}}{2000}]{Benitez00}
{Ben{\'\i}tez} N.,  2000, \mn@doi [\apj] {10.1086/308947}, \href
  {https://ui.adsabs.harvard.edu/abs/2000ApJ...536..571B} {536, 571}

\bibitem[\protect\citeauthoryear{{Bird}, {Viel}  \& {Haehnelt}}{{Bird}
  et~al.}{2012}]{Bird+12}
{Bird} S.,  {Viel} M.,   {Haehnelt} M.~G.,  2012, \mn@doi [\mnras]
  {10.1111/j.1365-2966.2011.20222.x}, \href
  {https://ui.adsabs.harvard.edu/\#abs/2012MNRAS.420.2551B} {420, 2551}

\bibitem[\protect\citeauthoryear{{Blandford}, {Saust}, {Brainerd}  \&
  {Villumsen}}{{Blandford} et~al.}{1991}]{Blandford+91}
{Blandford} R.~D.,  {Saust} A.~B.,  {Brainerd} T.~G.,   {Villumsen} J.~V.,
  1991, \mn@doi [\mnras] {10.1093/mnras/251.4.600}, \href
  {https://ui.adsabs.harvard.edu/abs/1991MNRAS.251..600B} {251, 600}

\bibitem[\protect\citeauthoryear{{Blazek}, {MacCrann}, {Troxel}  \&
  {Fang}}{{Blazek} et~al.}{2019}]{Blazek+19}
{Blazek} J.~A.,  {MacCrann} N.,  {Troxel} M.~A.,   {Fang} X.,  2019, \mn@doi
  [\prd] {10.1103/PhysRevD.100.103506}, \href
  {https://ui.adsabs.harvard.edu/abs/2019PhRvD.100j3506B} {100, 103506}

\bibitem[\protect\citeauthoryear{{Bocquet}, {Saro}, {Dolag}  \&
  {Mohr}}{{Bocquet} et~al.}{2016}]{Bocquet+16}
{Bocquet} S.,  {Saro} A.,  {Dolag} K.,   {Mohr} J.~J.,  2016, \mn@doi [\mnras]
  {10.1093/mnras/stv2657}, \href
  {https://ui.adsabs.harvard.edu/abs/2016MNRAS.456.2361B} {456, 2361}

\bibitem[\protect\citeauthoryear{{B{\"o}hringer}, {Chon}  \&
  {Fukugita}}{{B{\"o}hringer} et~al.}{2017}]{Boehringer+17}
{B{\"o}hringer} H.,  {Chon} G.,   {Fukugita} M.,  2017, \mn@doi [\aap]
  {10.1051/0004-6361/201731205}, \href
  {https://ui.adsabs.harvard.edu/abs/2017A&A...608A..65B} {608, A65}

\bibitem[\protect\citeauthoryear{{Bullock}, {Kolatt}, {Sigad}, {Somerville},
  {Kravtsov}, {Klypin}, {Primack}  \& {Dekel}}{{Bullock}
  et~al.}{2001}]{Bullock+01}
{Bullock} J.~S.,  {Kolatt} T.~S.,  {Sigad} Y.,  {Somerville} R.~S.,  {Kravtsov}
  A.~V.,  {Klypin} A.~A.,  {Primack} J.~R.,   {Dekel} A.,  2001, \mn@doi
  [\mnras] {10.1046/j.1365-8711.2001.04068.x}, \href
  {https://ui.adsabs.harvard.edu/abs/2001MNRAS.321..559B} {321, 559}

\bibitem[\protect\citeauthoryear{{Castro}, {Marra}  \& {Quartin}}{{Castro}
  et~al.}{2016}]{Castro+16}
{Castro} T.,  {Marra} V.,   {Quartin} M.,  2016, \mn@doi [\mnras]
  {10.1093/mnras/stw2072}, \href
  {https://ui.adsabs.harvard.edu/abs/2016MNRAS.463.1666C} {463, 1666}

\bibitem[\protect\citeauthoryear{{Castro}, {Borgani}, {Dolag}, {Marra},
  {Quartin}, {Saro}  \& {Sefusatti}}{{Castro} et~al.}{2021}]{Castro+21}
{Castro} T.,  {Borgani} S.,  {Dolag} K.,  {Marra} V.,  {Quartin} M.,  {Saro}
  A.,   {Sefusatti} E.,  2021, \mn@doi [\mnras] {10.1093/mnras/staa3473}, \href
  {https://ui.adsabs.harvard.edu/abs/2021MNRAS.500.2316C} {500, 2316}

\bibitem[\protect\citeauthoryear{{Chisari} et~al.,}{{Chisari}
  et~al.}{2018}]{Chisari+18}
{Chisari} N.~E.,  et~al., 2018, \mn@doi [\mnras] {10.1093/mnras/sty2093}, \href
  {https://ui.adsabs.harvard.edu/abs/2018MNRAS.480.3962C} {480, 3962}

\bibitem[\protect\citeauthoryear{{Chisari} et~al.,}{{Chisari}
  et~al.}{2019}]{Chisari+19}
{Chisari} N.~E.,  et~al., 2019, \mn@doi [\apjs] {10.3847/1538-4365/ab1658},
  \href {https://ui.adsabs.harvard.edu/abs/2019ApJS..242....2C} {242, 2}

\bibitem[\protect\citeauthoryear{{Costanzi}, {Villaescusa-Navarro}, {Viel},
  {Xia}, {Borgani}, {Castorina}  \& {Sefusatti}}{{Costanzi}
  et~al.}{2013}]{Costanzi+13}
{Costanzi} M.,  {Villaescusa-Navarro} F.,  {Viel} M.,  {Xia} J.-Q.,  {Borgani}
  S.,  {Castorina} E.,   {Sefusatti} E.,  2013, \mn@doi [\jcap]
  {10.1088/1475-7516/2013/12/012}, \href
  {https://ui.adsabs.harvard.edu/abs/2013JCAP...12..012C} {2013, 012}

\bibitem[\protect\citeauthoryear{{Costanzi} et~al.,}{{Costanzi}
  et~al.}{2019}]{Costanzi+19b}
{Costanzi} M.,  et~al., 2019, \mn@doi [\mnras] {10.1093/mnras/stz1949}, \href
  {https://ui.adsabs.harvard.edu/abs/2019MNRAS.488.4779C} {488, 4779}

\bibitem[\protect\citeauthoryear{{Crocce} \& {Scoccimarro}}{{Crocce} \&
  {Scoccimarro}}{2006}]{CrocceScoccimarro06}
{Crocce} M.,  {Scoccimarro} R.,  2006, \mn@doi [\prd]
  {10.1103/PhysRevD.73.063520}, \href
  {https://ui.adsabs.harvard.edu/abs/2006PhRvD..73f3520C} {73, 063520}

\bibitem[\protect\citeauthoryear{{Crocce}, {Castander}, {Gazta{\~n}aga},
  {Fosalba}  \& {Carretero}}{{Crocce} et~al.}{2015}]{Crocce+15}
{Crocce} M.,  {Castander} F.~J.,  {Gazta{\~n}aga} E.,  {Fosalba} P.,
  {Carretero} J.,  2015, \mn@doi [\mnras] {10.1093/mnras/stv1708}, \href
  {https://ui.adsabs.harvard.edu/abs/2015MNRAS.453.1513C} {453, 1513}

\bibitem[\protect\citeauthoryear{{DES Collaboration} et~al.,}{{DES
  Collaboration} et~al.}{2022}]{DES+22}
{DES Collaboration} et~al., 2022, \mn@doi [\prd] {10.1103/PhysRevD.105.023520},
  \href {https://ui.adsabs.harvard.edu/abs/2022PhRvD.105b3520A} {105, 023520}

\bibitem[\protect\citeauthoryear{{DES and KiDS collaboration} et~al.,}{{DES and
  KiDS collaboration} et~al.}{2023}]{DES+KiDS23}
{DES and KiDS collaboration} D.,  et~al., 2023, \mn@doi [arXiv e-prints]
  {10.48550/arXiv.2305.17173}, \href
  {https://ui.adsabs.harvard.edu/abs/2023arXiv230517173E} {p. arXiv:2305.17173}

\bibitem[\protect\citeauthoryear{{Despali}, {Giocoli}, {Angulo}, {Tormen},
  {Sheth}, {Baso}  \& {Moscardini}}{{Despali} et~al.}{2016}]{Despali+16}
{Despali} G.,  {Giocoli} C.,  {Angulo} R.~E.,  {Tormen} G.,  {Sheth} R.~K.,
  {Baso} G.,   {Moscardini} L.,  2016, \mn@doi [\mnras]
  {10.1093/mnras/stv2842}, \href
  {https://ui.adsabs.harvard.edu/abs/2016MNRAS.456.2486D} {456, 2486}

\bibitem[\protect\citeauthoryear{{Diemer}}{{Diemer}}{2021}]{Diemer21}
{Diemer} B.,  2021, \mn@doi [\apj] {10.3847/1538-4357/abd947}, \href
  {https://ui.adsabs.harvard.edu/abs/2021ApJ...909..112D} {909, 112}

\bibitem[\protect\citeauthoryear{{Drlica-Wagner} et~al.,}{{Drlica-Wagner}
  et~al.}{2018}]{Drlica-Wagner+18}
{Drlica-Wagner} A.,  et~al., 2018, \mn@doi [\apjs] {10.3847/1538-4365/aab4f5},
  \href {https://ui.adsabs.harvard.edu/abs/2018ApJS..235...33D} {235, 33}

\bibitem[\protect\citeauthoryear{{Edge}, {Sutherland}, {Kuijken}, {Driver},
  {McMahon}, {Eales}  \& {Emerson}}{{Edge} et~al.}{2013}]{Edge+13}
{Edge} A.,  {Sutherland} W.,  {Kuijken} K.,  {Driver} S.,  {McMahon} R.,
  {Eales} S.,   {Emerson} J.~P.,  2013, The Messenger, \href
  {https://ui.adsabs.harvard.edu/abs/2013Msngr.154...32E} {154, 32}

\bibitem[\protect\citeauthoryear{{Fang}, {Eifler}  \& {Krause}}{{Fang}
  et~al.}{2020}]{Fang+20}
{Fang} X.,  {Eifler} T.,   {Krause} E.,  2020, \mn@doi [\mnras]
  {10.1093/mnras/staa1726}, \href
  {https://ui.adsabs.harvard.edu/abs/2020MNRAS.497.2699F} {497, 2699}

\bibitem[\protect\citeauthoryear{{Feroz}, {Hobson}  \& {Bridges}}{{Feroz}
  et~al.}{2009}]{Multinest}
{Feroz} F.,  {Hobson} M.~P.,   {Bridges} M.,  2009, \mn@doi [\mnras]
  {10.1111/j.1365-2966.2009.14548.x}, \href
  {https://ui.adsabs.harvard.edu/abs/2009MNRAS.398.1601F} {398, 1601}

\bibitem[\protect\citeauthoryear{{Flaugher} et~al.,}{{Flaugher}
  et~al.}{2015}]{Flaugher+15}
{Flaugher} B.,  et~al., 2015, \mn@doi [\aj] {10.1088/0004-6256/150/5/150},
  \href {https://ui.adsabs.harvard.edu/abs/2015AJ....150..150F} {150, 150}

\bibitem[\protect\citeauthoryear{{Fosalba}, {Gazta{\~n}aga}, {Castander}  \&
  {Crocce}}{{Fosalba} et~al.}{2015a}]{Fosalba+15b}
{Fosalba} P.,  {Gazta{\~n}aga} E.,  {Castander} F.~J.,   {Crocce} M.,  2015a,
  \mn@doi [\mnras] {10.1093/mnras/stu2464}, \href
  {https://ui.adsabs.harvard.edu/abs/2015MNRAS.447.1319F} {447, 1319}

\bibitem[\protect\citeauthoryear{{Fosalba}, {Crocce}, {Gazta{\~n}aga}  \&
  {Castander}}{{Fosalba} et~al.}{2015b}]{Fosalba+15a}
{Fosalba} P.,  {Crocce} M.,  {Gazta{\~n}aga} E.,   {Castander} F.~J.,  2015b,
  \mn@doi [\mnras] {10.1093/mnras/stv138}, \href
  {https://ui.adsabs.harvard.edu/abs/2015MNRAS.448.2987F} {448, 2987}

\bibitem[\protect\citeauthoryear{{Gatti} et~al.,}{{Gatti}
  et~al.}{2022}]{Gatti+22}
{Gatti} M.,  et~al., 2022, \mn@doi [\mnras] {10.1093/mnras/stab3311}, \href
  {https://ui.adsabs.harvard.edu/abs/2022MNRAS.510.1223G} {510, 1223}

\bibitem[\protect\citeauthoryear{{Geach}}{{Geach}}{2012}]{Geach12}
{Geach} J.~E.,  2012, \mn@doi [\mnras] {10.1111/j.1365-2966.2011.19913.x},
  \href {https://ui.adsabs.harvard.edu/abs/2012MNRAS.419.2633G} {419, 2633}

\bibitem[\protect\citeauthoryear{{Giblin} et~al.,}{{Giblin}
  et~al.}{2021}]{Giblin+21}
{Giblin} B.,  et~al., 2021, \mn@doi [\aap] {10.1051/0004-6361/202038850}, \href
  {https://ui.adsabs.harvard.edu/abs/2021A&A...645A.105G} {645, A105}

\bibitem[\protect\citeauthoryear{{Gunn}}{{Gunn}}{1967}]{Gunn67}
{Gunn} J.~E.,  1967, \mn@doi [\apj] {10.1086/149378}, \href
  {https://ui.adsabs.harvard.edu/abs/1967ApJ...150..737G} {150, 737}

\bibitem[\protect\citeauthoryear{{Hamilton}}{{Hamilton}}{2000}]{Hamilton00}
{Hamilton} A.~J.~S.,  2000, \mn@doi [\mnras]
  {10.1046/j.1365-8711.2000.03071.x}, \href
  {https://ui.adsabs.harvard.edu/abs/2000MNRAS.312..257H} {312, 257}

\bibitem[\protect\citeauthoryear{{Harnois-D{\'e}raps} \& {van
  Waerbeke}}{{Harnois-D{\'e}raps} \& {van Waerbeke}}{2015}]{HDvW15}
{Harnois-D{\'e}raps} J.,  {van Waerbeke} L.,  2015, \mn@doi [\mnras]
  {10.1093/mnras/stv794}, \href
  {https://ui.adsabs.harvard.edu/abs/2015MNRAS.450.2857H} {450, 2857}

\bibitem[\protect\citeauthoryear{{Harnois-D{\'e}raps}
  et~al.,}{{Harnois-D{\'e}raps} et~al.}{2018}]{Harnois-Deraps+18}
{Harnois-D{\'e}raps} J.,  et~al., 2018, \mn@doi [\mnras]
  {10.1093/mnras/sty2319}, \href
  {https://ui.adsabs.harvard.edu/abs/2018MNRAS.481.1337H} {481, 1337}

\bibitem[\protect\citeauthoryear{Harris et~al.,}{Harris et~al.}{2020}]{numpy}
Harris C.~R.,  et~al., 2020, \mn@doi [Nature] {10.1038/s41586-020-2649-2}, 585,
  357

\bibitem[\protect\citeauthoryear{{Hartley} et~al.,}{{Hartley}
  et~al.}{2022}]{Hartley+22}
{Hartley} W.~G.,  et~al., 2022, \mn@doi [\mnras] {10.1093/mnras/stab3055},
  \href {https://ui.adsabs.harvard.edu/abs/2022MNRAS.509.3547H} {509, 3547}

\bibitem[\protect\citeauthoryear{{Heymans} et~al.,}{{Heymans}
  et~al.}{2012}]{CFHTLS}
{Heymans} C.,  et~al., 2012, \mn@doi [\mnras]
  {10.1111/j.1365-2966.2012.21952.x}, \href
  {https://ui.adsabs.harvard.edu/\#abs/2012MNRAS.427..146H} {427, 146}

\bibitem[\protect\citeauthoryear{{Heymans} et~al.,}{{Heymans}
  et~al.}{2021}]{Heymans+21}
{Heymans} C.,  et~al., 2021, \mn@doi [\aap] {10.1051/0004-6361/202039063},
  \href {https://ui.adsabs.harvard.edu/abs/2021A&A...646A.140H} {646, A140}

\bibitem[\protect\citeauthoryear{{Hildebrandt} et~al.,}{{Hildebrandt}
  et~al.}{2017}]{Hildebrandt+17}
{Hildebrandt} H.,  et~al., 2017, \mn@doi [\mnras] {10.1093/mnras/stw2805},
  \href {https://ui.adsabs.harvard.edu/\#abs/2017MNRAS.465.1454H} {465, 1454}

\bibitem[\protect\citeauthoryear{{Hildebrandt} et~al.,}{{Hildebrandt}
  et~al.}{2020}]{Hildebrandt+20}
{Hildebrandt} H.,  et~al., 2020, \mn@doi [\aap] {10.1051/0004-6361/201834878},
  \href {https://ui.adsabs.harvard.edu/abs/2020A&A...633A..69H} {633, A69}

\bibitem[\protect\citeauthoryear{{Hinshaw} et~al.,}{{Hinshaw}
  et~al.}{2013}]{WMAP9}
{Hinshaw} G.,  et~al., 2013, \mn@doi [\apjs] {10.1088/0067-0049/208/2/19},
  \href {http://adsabs.harvard.edu/abs/2013ApJS..208...19H} {208, 19}

\bibitem[\protect\citeauthoryear{{Hoekstra}, {Donahue}, {Conselice}, {McNamara}
   \& {Voit}}{{Hoekstra} et~al.}{2011}]{Hoekstra+11}
{Hoekstra} H.,  {Donahue} M.,  {Conselice} C.~J.,  {McNamara} B.~R.,   {Voit}
  G.~M.,  2011, \mn@doi [\apj] {10.1088/0004-637X/726/1/48}, \href
  {https://ui.adsabs.harvard.edu/abs/2011ApJ...726...48H} {726, 48}

\bibitem[\protect\citeauthoryear{{Hubble}}{{Hubble}}{1936}]{Hubble36b}
{Hubble} E.~P.,  1936, {Realm of the Nebulae}

\bibitem[\protect\citeauthoryear{Hunter}{Hunter}{2007}]{matplotlib}
Hunter J.~D.,  2007, \mn@doi [Computing in Science \& Engineering]
  {10.1109/MCSE.2007.55}, 9, 90

\bibitem[\protect\citeauthoryear{{Jain} \& {Taylor}}{{Jain} \&
  {Taylor}}{2003}]{JainTaylor03}
{Jain} B.,  {Taylor} A.,  2003, \mn@doi [\prl] {10.1103/PhysRevLett.91.141302},
  \href {https://ui.adsabs.harvard.edu/abs/2003PhRvL..91n1302J} {91, 141302}

\bibitem[\protect\citeauthoryear{{Jenkins}, {Frenk}, {White}, {Colberg},
  {Cole}, {Evrard}, {Couchman}  \& {Yoshida}}{{Jenkins}
  et~al.}{2001}]{jenkins01}
{Jenkins} A.,  {Frenk} C.~S.,  {White} S.~D.~M.,  {Colberg} J.~M.,  {Cole} S.,
  {Evrard} A.~E.,  {Couchman} H.~M.~P.,   {Yoshida} N.,  2001, \mn@doi [\mnras]
  {10.1046/j.1365-8711.2001.04029.x}, \href
  {http://adsabs.harvard.edu/abs/2001MNRAS.321..372J} {321, 372}

\bibitem[\protect\citeauthoryear{{Joachimi} et~al.,}{{Joachimi}
  et~al.}{2020}]{Joachimi+20}
{Joachimi} B.,  et~al., 2020, arXiv e-prints, \href
  {https://ui.adsabs.harvard.edu/abs/2020arXiv200701844J} {p. arXiv:2007.01844}

\bibitem[\protect\citeauthoryear{{Knebe} et~al.,}{{Knebe}
  et~al.}{2011}]{Knebe+11}
{Knebe} A.,  et~al., 2011, \mn@doi [\mnras] {10.1111/j.1365-2966.2011.18858.x},
  \href {https://ui.adsabs.harvard.edu/abs/2011MNRAS.415.2293K} {415, 2293}

\bibitem[\protect\citeauthoryear{{Kohonen}}{{Kohonen}}{2001}]{SOM}
{Kohonen} T.,  2001, {Self-Organizing Maps}.
Springer series in information sciences, 2001, xx, 501

\bibitem[\protect\citeauthoryear{{Krause} \& {Eifler}}{{Krause} \&
  {Eifler}}{2017}]{KrauseEifler17}
{Krause} E.,  {Eifler} T.,  2017, \mn@doi [\mnras] {10.1093/mnras/stx1261},
  \href {https://ui.adsabs.harvard.edu/abs/2017MNRAS.470.2100K} {470, 2100}

\bibitem[\protect\citeauthoryear{{Kuijken}}{{Kuijken}}{2011}]{Kuijken11}
{Kuijken} K.,  2011, The Messenger, \href
  {https://ui.adsabs.harvard.edu/abs/2011Msngr.146....8K} {146, 8}

\bibitem[\protect\citeauthoryear{{Kuijken} et~al.,}{{Kuijken}
  et~al.}{2015}]{Kuijken+15}
{Kuijken} K.,  et~al., 2015, \mn@doi [\mnras] {10.1093/mnras/stv2140}, \href
  {https://ui.adsabs.harvard.edu/abs/2015MNRAS.454.3500K} {454, 3500}

\bibitem[\protect\citeauthoryear{{Kuijken} et~al.,}{{Kuijken}
  et~al.}{2019}]{Kuijken+19}
{Kuijken} K.,  et~al., 2019, \mn@doi [\aap] {10.1051/0004-6361/201834918},
  \href {https://ui.adsabs.harvard.edu/abs/2019A&A...625A...2K} {625, A2}

\bibitem[\protect\citeauthoryear{{Kulkarni} \& {Ostriker}}{{Kulkarni} \&
  {Ostriker}}{2022}]{Kulkarni+22}
{Kulkarni} M.,  {Ostriker} J.~P.,  2022, \mn@doi [\mnras]
  {10.1093/mnras/stab3520}, \href
  {https://ui.adsabs.harvard.edu/abs/2022MNRAS.510.1425K} {510, 1425}

\bibitem[\protect\citeauthoryear{{Leauthaud} et~al.,}{{Leauthaud}
  et~al.}{2017}]{Leauthaud+17}
{Leauthaud} A.,  et~al., 2017, \mn@doi [\mnras] {10.1093/mnras/stx258}, \href
  {https://ui.adsabs.harvard.edu/abs/2017MNRAS.467.3024L} {467, 3024}

\bibitem[\protect\citeauthoryear{{Lemos} et~al.,}{{Lemos}
  et~al.}{2022}]{Lemos+22}
{Lemos} P.,  et~al., 2022, \mn@doi [\mnras] {10.1093/mnras/stac2786}, \href
  {https://ui.adsabs.harvard.edu/abs/2022MNRAS.tmp.2714L} {}

\bibitem[\protect\citeauthoryear{{Lewis}}{{Lewis}}{2019}]{getdist}
{Lewis} A.,  2019, arXiv e-prints, \href
  {https://ui.adsabs.harvard.edu/abs/2019arXiv191013970L} {p. arXiv:1910.13970}

\bibitem[\protect\citeauthoryear{{Lewis} \& {Challinor}}{{Lewis} \&
  {Challinor}}{2011}]{CAMB}
{Lewis} A.,  {Challinor} A.,  2011, {CAMB: Code for Anisotropies in the
  Microwave Background}, Astrophysics Source Code Library, record ascl:1102.026
  (\mn@eprint {ascl} {1102.026})

\bibitem[\protect\citeauthoryear{{Li}, {Lelli}, {McGaugh}, {Pawlowski}, {Zwaan}
   \& {Schombert}}{{Li} et~al.}{2019}]{Li+19}
{Li} P.,  {Lelli} F.,  {McGaugh} S.,  {Pawlowski} M.~S.,  {Zwaan} M.~A.,
  {Schombert} J.,  2019, \mn@doi [\apjl] {10.3847/2041-8213/ab53e6}, \href
  {https://ui.adsabs.harvard.edu/abs/2019ApJ...886L..11L} {886, L11}

\bibitem[\protect\citeauthoryear{{Limber}}{{Limber}}{1953}]{Limber53}
{Limber} D.~N.,  1953, \mn@doi [\apj] {10.1086/145672}, \href
  {https://ui.adsabs.harvard.edu/abs/1953ApJ...117..134L} {117, 134}

\bibitem[\protect\citeauthoryear{{Lovell}}{{Lovell}}{2020}]{Lovell20}
{Lovell} M.~R.,  2020, \mn@doi [\mnras] {10.1093/mnrasl/slaa005}, \href
  {https://ui.adsabs.harvard.edu/abs/2020MNRAS.493L..11L} {493, L11}

\bibitem[\protect\citeauthoryear{{Marsh}}{{Marsh}}{2015}]{Marsh15}
{Marsh} D. J.~E.,  2015, \mn@doi [\prd] {10.1103/PhysRevD.91.123520}, \href
  {https://ui.adsabs.harvard.edu/abs/2015PhRvD..91l3520M} {91, 123520}

\bibitem[\protect\citeauthoryear{{Mead}, {Peacock}, {Heymans}, {Joudaki}  \&
  {Heavens}}{{Mead} et~al.}{2015}]{Mead+15}
{Mead} A.~J.,  {Peacock} J.~A.,  {Heymans} C.,  {Joudaki} S.,   {Heavens}
  A.~F.,  2015, \mn@doi [\mnras] {10.1093/mnras/stv2036}, \href
  {https://ui.adsabs.harvard.edu/abs/2015MNRAS.454.1958M} {454, 1958}

\bibitem[\protect\citeauthoryear{{Mead}, {Heymans}, {Lombriser}, {Peacock},
  {Steele}  \& {Winther}}{{Mead} et~al.}{2016}]{Mead+16}
{Mead} A.~J.,  {Heymans} C.,  {Lombriser} L.,  {Peacock} J.~A.,  {Steele}
  O.~I.,   {Winther} H.~A.,  2016, \mn@doi [\mnras] {10.1093/mnras/stw681},
  \href {https://ui.adsabs.harvard.edu/abs/2016MNRAS.459.1468M} {459, 1468}

\bibitem[\protect\citeauthoryear{{Mead}, {Tr{\"o}ster}, {Heymans}, {Van
  Waerbeke}  \& {McCarthy}}{{Mead} et~al.}{2020}]{Mead+20a}
{Mead} A.~J.,  {Tr{\"o}ster} T.,  {Heymans} C.,  {Van Waerbeke} L.,
  {McCarthy} I.~G.,  2020, \mn@doi [\aap] {10.1051/0004-6361/202038308}, \href
  {https://ui.adsabs.harvard.edu/abs/2020A&A...641A.130M} {641, A130}

\bibitem[\protect\citeauthoryear{{Mead}, {Brieden}, {Tr{\"o}ster}  \&
  {Heymans}}{{Mead} et~al.}{2021}]{Mead+21}
{Mead} A.~J.,  {Brieden} S.,  {Tr{\"o}ster} T.,   {Heymans} C.,  2021, \mn@doi
  [\mnras] {10.1093/mnras/stab082}, \href
  {https://ui.adsabs.harvard.edu/abs/2021MNRAS.502.1401M} {502, 1401}

\bibitem[\protect\citeauthoryear{{Myles} et~al.,}{{Myles}
  et~al.}{2021}]{Myles+21}
{Myles} J.,  et~al., 2021, \mn@doi [\mnras] {10.1093/mnras/stab1515}, \href
  {https://ui.adsabs.harvard.edu/abs/2021MNRAS.505.4249M} {505, 4249}

\bibitem[\protect\citeauthoryear{{Newman}}{{Newman}}{2008}]{Newman08}
{Newman} J.~A.,  2008, \mn@doi [\apj] {10.1086/589982}, \href
  {https://ui.adsabs.harvard.edu/abs/2008ApJ...684...88N} {684, 88}

\bibitem[\protect\citeauthoryear{{Ondaro-Mallea}, {Angulo}, {Zennaro},
  {Contreras}  \& {Aric{\`o}}}{{Ondaro-Mallea} et~al.}{2022}]{Ondaro-Mallea+22}
{Ondaro-Mallea} L.,  {Angulo} R.~E.,  {Zennaro} M.,  {Contreras} S.,
  {Aric{\`o}} G.,  2022, \mn@doi [\mnras] {10.1093/mnras/stab3337}, \href
  {https://ui.adsabs.harvard.edu/abs/2022MNRAS.509.6077O} {509, 6077}

\bibitem[\protect\citeauthoryear{{Peacock} \& {Smith}}{{Peacock} \&
  {Smith}}{2000}]{PeacockSmith00}
{Peacock} J.~A.,  {Smith} R.~E.,  2000, \mn@doi [\mnras]
  {10.1046/j.1365-8711.2000.03779.x}, \href
  {http://adsabs.harvard.edu/abs/2000MNRAS.318.1144P} {318, 1144}

\bibitem[\protect\citeauthoryear{{Planck Collaboration} et~al.,}{{Planck
  Collaboration} et~al.}{2020}]{planck18}
{Planck Collaboration} et~al., 2020, \mn@doi [\aap]
  {10.1051/0004-6361/201833910}, \href
  {https://ui.adsabs.harvard.edu/abs/2020A&A...641A...6P} {641, A6}

\bibitem[\protect\citeauthoryear{{Press} \& {Schechter}}{{Press} \&
  {Schechter}}{1974}]{PSHMF}
{Press} W.~H.,  {Schechter} P.,  1974, \mn@doi [\apj] {10.1086/152650}, \href
  {https://ui.adsabs.harvard.edu/\#abs/1974ApJ...187..425P} {187, 425}

\bibitem[\protect\citeauthoryear{{Raichoor} et~al.,}{{Raichoor}
  et~al.}{2014}]{Raichoor+14}
{Raichoor} A.,  et~al., 2014, \mn@doi [\apj] {10.1088/0004-637X/797/2/102},
  \href {https://ui.adsabs.harvard.edu/abs/2014ApJ...797..102R} {797, 102}

\bibitem[\protect\citeauthoryear{{S{\'a}nchez} et~al.,}{{S{\'a}nchez}
  et~al.}{2022}]{Sanchez+22}
{S{\'a}nchez} C.,  et~al., 2022, \mn@doi [\prd] {10.1103/PhysRevD.105.083529},
  \href {https://ui.adsabs.harvard.edu/abs/2022PhRvD.105h3529S} {105, 083529}

\bibitem[\protect\citeauthoryear{{Schaye} et~al.,}{{Schaye}
  et~al.}{2023}]{Schaye+23}
{Schaye} J.,  et~al., 2023, \mn@doi [arXiv e-prints]
  {10.48550/arXiv.2306.04024}, \href
  {https://ui.adsabs.harvard.edu/abs/2023arXiv230604024S} {p. arXiv:2306.04024}

\bibitem[\protect\citeauthoryear{{Schneider} \& {Teyssier}}{{Schneider} \&
  {Teyssier}}{2015}]{SchneiderTeyssier15}
{Schneider} A.,  {Teyssier} R.,  2015, \mn@doi [\jcap]
  {10.1088/1475-7516/2015/12/049}, \href
  {https://ui.adsabs.harvard.edu/abs/2015JCAP...12..049S} {2015, 049}

\bibitem[\protect\citeauthoryear{{Schneider}, {Stoira}, {Refregier}, {Weiss},
  {Knabenhans}, {Stadel}  \& {Teyssier}}{{Schneider}
  et~al.}{2020}]{Schneider+20}
{Schneider} A.,  {Stoira} N.,  {Refregier} A.,  {Weiss} A.~J.,  {Knabenhans}
  M.,  {Stadel} J.,   {Teyssier} R.,  2020, \mn@doi [\jcap]
  {10.1088/1475-7516/2020/04/019}, \href
  {https://ui.adsabs.harvard.edu/abs/2020JCAP...04..019S} {2020, 019}

\bibitem[\protect\citeauthoryear{{Secco} et~al.,}{{Secco}
  et~al.}{2022}]{Secco+22}
{Secco} L.~F.,  et~al., 2022, \mn@doi [\prd] {10.1103/PhysRevD.105.023515},
  \href {https://ui.adsabs.harvard.edu/abs/2022PhRvD.105b3515S} {105, 023515}

\bibitem[\protect\citeauthoryear{{Sevilla-Noarbe} et~al.,}{{Sevilla-Noarbe}
  et~al.}{2021}]{Sevilla-Noarbe+21}
{Sevilla-Noarbe} I.,  et~al., 2021, \mn@doi [\apjs] {10.3847/1538-4365/abeb66},
  \href {https://ui.adsabs.harvard.edu/abs/2021ApJS..254...24S} {254, 24}

\bibitem[\protect\citeauthoryear{{Sheth} \& {Tormen}}{{Sheth} \&
  {Tormen}}{1999}]{STHMF}
{Sheth} R.~K.,  {Tormen} G.,  1999, \mn@doi [\mnras]
  {10.1046/j.1365-8711.1999.02692.x}, \href
  {https://ui.adsabs.harvard.edu/abs/1999MNRAS.308..119S} {308, 119}

\bibitem[\protect\citeauthoryear{{Sheth}, {Mo}  \& {Tormen}}{{Sheth}
  et~al.}{2001}]{Sheth+01}
{Sheth} R.~K.,  {Mo} H.~J.,   {Tormen} G.,  2001, \mn@doi [\mnras]
  {10.1046/j.1365-8711.2001.04006.x}, \href
  {https://ui.adsabs.harvard.edu/abs/2001MNRAS.323....1S} {323, 1}

\bibitem[\protect\citeauthoryear{{Smith} et~al.,}{{Smith}
  et~al.}{2003}]{HALOFIT}
{Smith} R.~E.,  et~al., 2003, \mn@doi [\mnras]
  {10.1046/j.1365-8711.2003.06503.x}, \href
  {https://ui.adsabs.harvard.edu/\#abs/2003MNRAS.341.1311S} {341, 1311}

\bibitem[\protect\citeauthoryear{{Smith}, {Scoccimarro}  \& {Sheth}}{{Smith}
  et~al.}{2007}]{Smith+07}
{Smith} R.~E.,  {Scoccimarro} R.,   {Sheth} R.~K.,  2007, \mn@doi [\prd]
  {10.1103/PhysRevD.75.063512}, \href
  {https://ui.adsabs.harvard.edu/abs/2007PhRvD..75f3512S} {75, 063512}

\bibitem[\protect\citeauthoryear{{Takahashi}, {Sato}, {Nishimichi}, {Taruya}
  \& {Oguri}}{{Takahashi} et~al.}{2012}]{Takahashi+12}
{Takahashi} R.,  {Sato} M.,  {Nishimichi} T.,  {Taruya} A.,   {Oguri} M.,
  2012, \mn@doi [\apj] {10.1088/0004-637X/761/2/152}, \href
  {https://ui.adsabs.harvard.edu/\#abs/2012ApJ...761..152T} {761, 152}

\bibitem[\protect\citeauthoryear{{Tormen}}{{Tormen}}{1998}]{Tormen98}
{Tormen} G.,  1998, \mn@doi [\mnras] {10.1046/j.1365-8711.1998.01545.x}, \href
  {https://ui.adsabs.harvard.edu/abs/1998MNRAS.297..648T} {297, 648}

\bibitem[\protect\citeauthoryear{{Tr{\"o}ster} et~al.,}{{Tr{\"o}ster}
  et~al.}{2021}]{Troester+21}
{Tr{\"o}ster} T.,  et~al., 2021, \mn@doi [\aap] {10.1051/0004-6361/202039805},
  \href {https://ui.adsabs.harvard.edu/abs/2021A&A...649A..88T} {649, A88}

\bibitem[\protect\citeauthoryear{{Tr{\"o}ster} et~al.,}{{Tr{\"o}ster}
  et~al.}{2022}]{Troester+22}
{Tr{\"o}ster} T.,  et~al., 2022, \mn@doi [\aap] {10.1051/0004-6361/202142197},
  \href {https://ui.adsabs.harvard.edu/abs/2022A&A...660A..27T} {660, A27}

\bibitem[\protect\citeauthoryear{Virtanen et~al.,}{Virtanen
  et~al.}{2020}]{scipy}
Virtanen P.,  et~al., 2020, \mn@doi [Nature Methods]
  {10.1038/s41592-019-0686-2}, \href {https://rdcu.be/b08Wh} {17, 261}

\bibitem[\protect\citeauthoryear{{White} \& {Frenk}}{{White} \&
  {Frenk}}{1991}]{WFSAM}
{White} S.~D.~M.,  {Frenk} C.~S.,  1991, \mn@doi [\apj] {10.1086/170483}, \href
  {http://adsabs.harvard.edu/abs/1991ApJ...379...52W} {379, 52}

\bibitem[\protect\citeauthoryear{{Wright}, {Hildebrandt}, {van den Busch},
  {Heymans}, {Joachimi}, {Kannawadi}  \& {Kuijken}}{{Wright}
  et~al.}{2020}]{Wright+20b}
{Wright} A.~H.,  {Hildebrandt} H.,  {van den Busch} J.~L.,  {Heymans} C.,
  {Joachimi} B.,  {Kannawadi} A.,   {Kuijken} K.,  2020, \mn@doi [\aap]
  {10.1051/0004-6361/202038389}, \href
  {https://ui.adsabs.harvard.edu/abs/2020A&A...640L..14W} {640, L14}

\bibitem[\protect\citeauthoryear{{Zuntz} et~al.,}{{Zuntz}
  et~al.}{2015}]{Zuntz+15}
{Zuntz} J.,  et~al., 2015, \mn@doi [Astronomy and Computing]
  {10.1016/j.ascom.2015.05.005}, \href
  {https://ui.adsabs.harvard.edu/abs/2015A&C....12...45Z} {12, 45}

\bibitem[\protect\citeauthoryear{{de Jong} et~al.,}{{de Jong}
  et~al.}{2017}]{deJong+17}
{de Jong} J. T.~A.,  et~al., 2017, \mn@doi [\aap]
  {10.1051/0004-6361/201730747}, \href
  {https://ui.adsabs.harvard.edu/abs/2017A&A...604A.134D} {604, A134}

\bibitem[\protect\citeauthoryear{{eBOSS Collaboration} et~al.,}{{eBOSS
  Collaboration} et~al.}{2020}]{eboss20}
{eBOSS Collaboration} et~al., 2020, arXiv e-prints, \href
  {https://ui.adsabs.harvard.edu/abs/2020arXiv200708991E} {p. arXiv:2007.08991}

\bibitem[\protect\citeauthoryear{{van Daalen}, {McCarthy}  \& {Schaye}}{{van
  Daalen} et~al.}{2020}]{VanDaalen+20}
{van Daalen} M.~P.,  {McCarthy} I.~G.,   {Schaye} J.,  2020, \mn@doi [\mnras]
  {10.1093/mnras/stz3199}, \href
  {https://ui.adsabs.harvard.edu/abs/2020MNRAS.491.2424V} {491, 2424}

\bibitem[\protect\citeauthoryear{{van den Busch} et~al.,}{{van den Busch}
  et~al.}{2022}]{vandenBusch+22}
{van den Busch} J.~L.,  et~al., 2022, \mn@doi [\aap]
  {10.1051/0004-6361/202142083}, \href
  {https://ui.adsabs.harvard.edu/abs/2022A&A...664A.170V} {664, A170}

\makeatother
\end{thebibliography}




\appendix

\section{Sampling and Scale Cuts}
\label{asecsc}

\subsection{Impact of Sampling}

\begin{figure}
\centering
\includegraphics[width=0.45\textwidth]{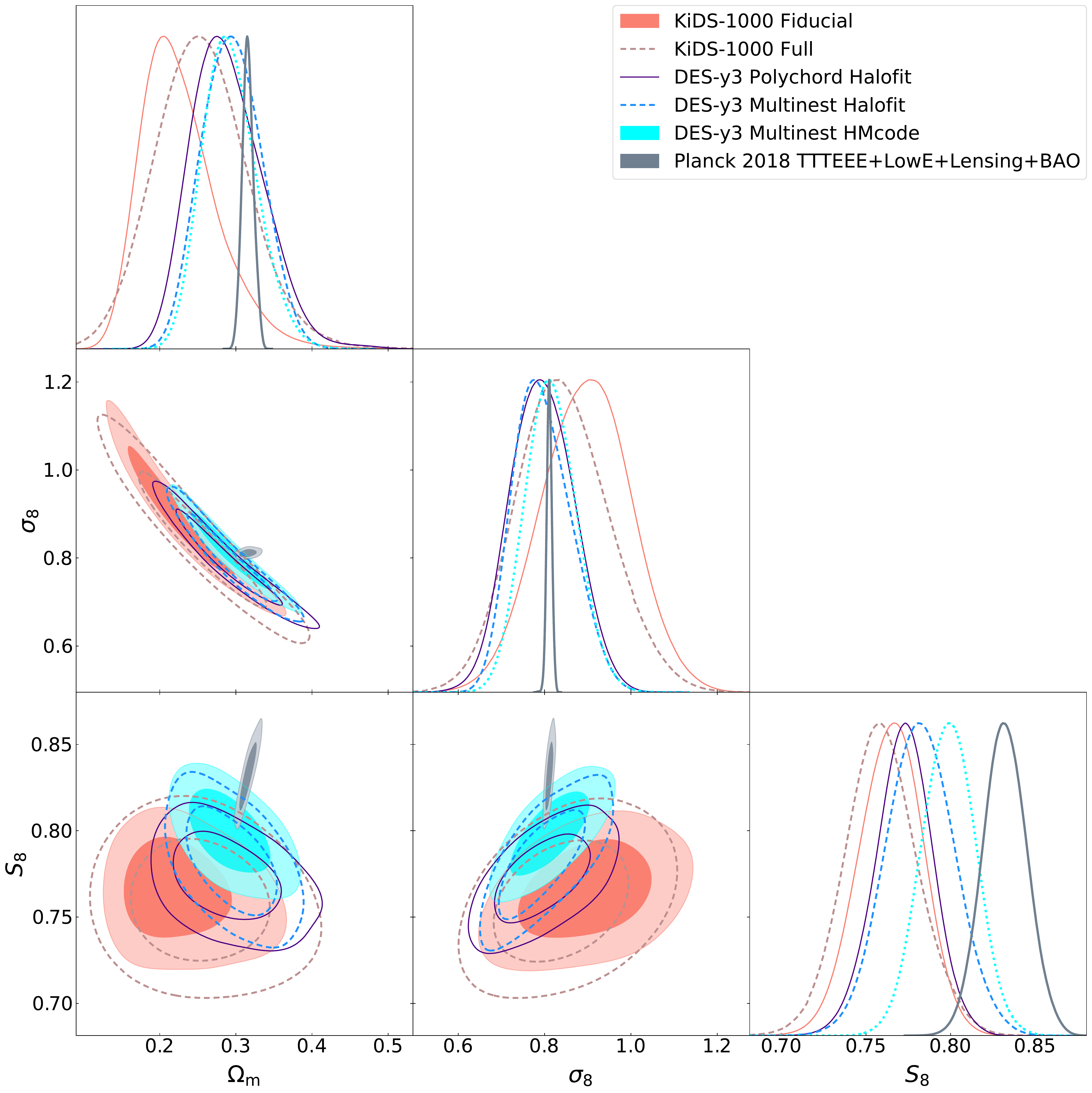}
\caption{Posterior contours ($68\%$ and $95\%$) in the $\Om-\se$, $\Om-\Se$ and $\se-\Se$ planes for KiDS-1000, DES-y3, and {\it Planck} 2018. The filled \salmon{salmon pink} contours (KiDS-1000 Fiducial) are the $\xi_\pm$ analysis using the fiducial scale cut from \citet{Asgari+21}. The dashed \rosybrown{rosybrown} contours (KiDS-1000 Full) use all data points from KiDS-1000 $\xi_\pm$. The solid \indigo{indigo blue} and the dashed \brightazure{bright azure} contours are the \lcdm-optimised DES-Y3 cosmic shear results with \textsc{Halofit}, using \indigo{\textsc{polychord}} (fiducial) and \brightazure{\textsc{multinest}} respectively. The filled \cyan{cyan} contours are the \textsc{HMcode}-based DES-y3 results under KiDS-1000 prior of $B$ parameters. The \slategrey{slate grey} contours are the TTTEEE+low$\ell$+low E+lensing+BAO result from \citet{planck18} as the comparison to lensing results.
}\label{fig:scale_cut_0}
\end{figure}

Here, we show how a particular choice of sampling may affect the posterior distribution in cosmic shear calculations.
Figure \ref{fig:scale_cut_0} shows different combinations of sampling (\textsc{multinest} versus \textsc{polychord}) and non-linear power spectrum calculations (\textsc{Halofit} versus \textsc{HMcode}). The original DES-y3 setup, \lcdm-optimised cosmic shear results, which uses \textsc{polychord} with \textsc{Halofit}, is shown by the indigo contours. It shows a $2.8\sigma$ tension in $\Se$ compared to the \citet{planck18} cosmology. The \textsc{multinest} sampling is shown by the cyan dashed contours, and finally, the \textsc{multinest} using \textsc{HMcode} is shown by the cyan solid contours. The latter has only a $1.6\sigma$ tension with the {\it Planck} value. When we use \textsc{HMcode}, we always set the halo concentration parameter $B$ from \citet{Bullock+01} as a free parameter ($B\in[2,3.13]$) -- like the set-up of the KiDS-1000 fiducial analysis from \citet{Asgari+21}.

\subsection{Impact of scale cuts}

\begin{figure}
\centering
\includegraphics[width=0.47\textwidth]{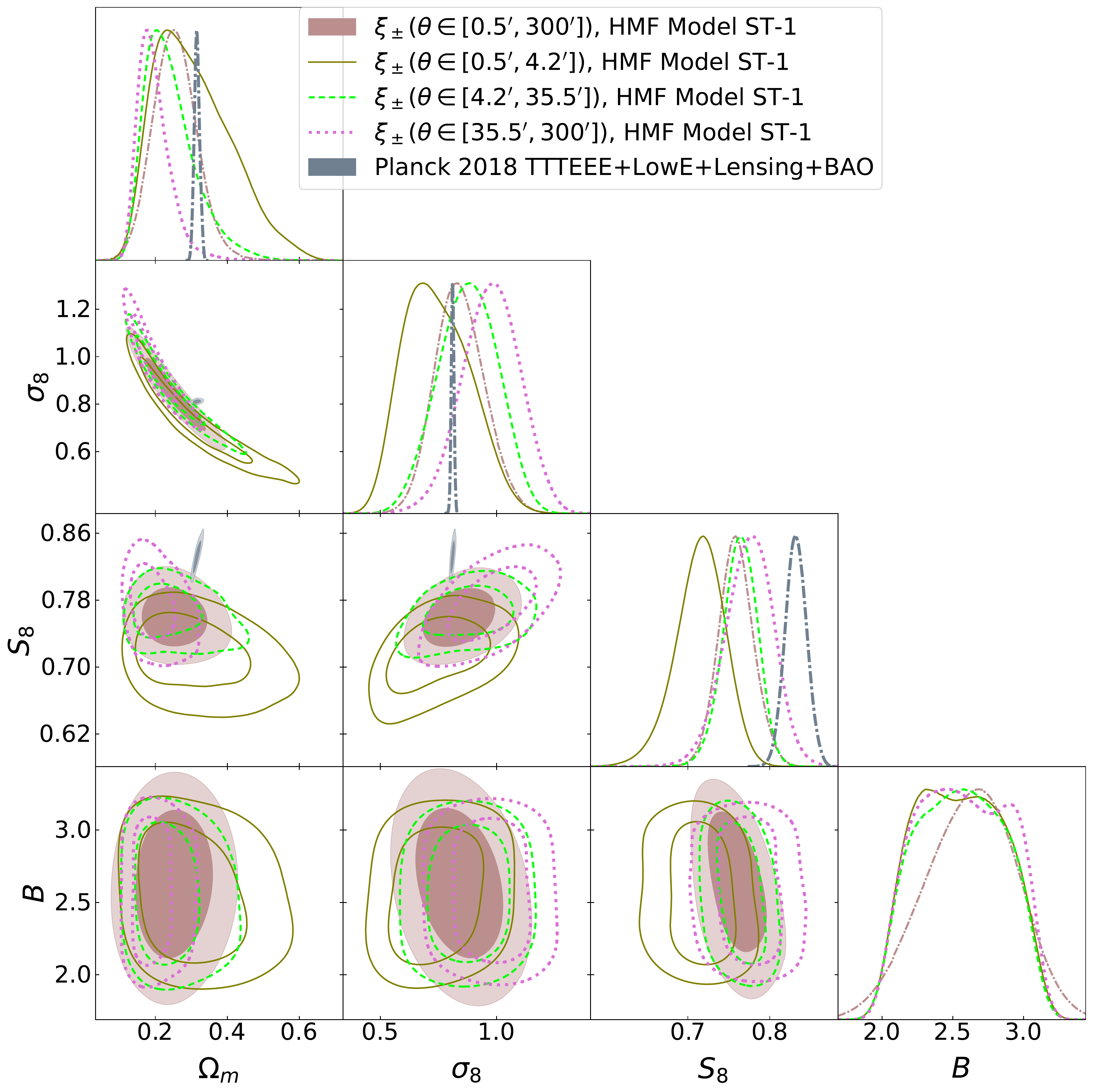}
\caption{Posterior contours ($68\%$ and $95\%$) in the $\Om-\se-\Se-B$ parameter space for KiDS-1000 given the Standard Sheth-Tormen HMF for different scale cuts. The \rosybrown{rosybrown} solid contour is for no scale cut ($\theta \in (0.5',300')$). The \olive{olive} solid line, \lime{lime} dashed line and \orchid{orchid} dotted line contours are for the scale cuts bins $\theta \in (0.5',4.2')$, $\theta \in (4.2',35.5')$ and $\theta \in (35.5',300')$, respectively.}
\label{fig:scale_cut_cosmo1}
\end{figure}

Different choices of scale cuts may also impact the posterior distributions by altering the sensitivity of the statistical estimator to various physical effects. For instance, it is known that small scales are more sensitive to baryonic physics.
The pink solid contours and the brown dashed contours in Figure \ref{fig:scale_cut_0} respectively show the fiducial KiDS-1000 analysis (also appeared as ST-1 cosmological analysis in our main text) and the analysis with no scale cuts at all ($\theta \in [0.5',300']$ for both $\xi_\pm$). The KiDS-1000 fiducial analysis applied a $\theta > 4'$ cut. The difference between the two is marginal.

Figure \ref{fig:scale_cut_cosmo1} shows the result of various scale cuts on KiDS-1000 data, assuming the HMF parameters are fixed to ST1. We use three bins of different scale-cut, separated by the geometric progressed values of $4.2'$ and $35.5'$. We can see that the different scale bins do not show a particular trend for a low or high $\Se$, which means that different choices of scale cuts do not impact the cosmological inference significantly, at least for this data set and survey area.

\bsp	
\label{lastpage}
\end{document}